\documentclass[a4paper,12pt]{article}
\usepackage{amsmath}
\usepackage{accents}
\usepackage{bm}
\usepackage{dsfont}
\usepackage[english]{babel}
\usepackage{graphicx}
\usepackage[usenames]{color}
\usepackage{colortbl}
\usepackage{mathtools}
\usepackage{commath}
\usepackage[font=footnotesize,figurewithin=none]{caption}
\usepackage[caption=false]{subfig}
\usepackage[toc,page,title]{appendix}

\begin{document}

\centerline {\LARGE{Entanglement of the Ising-Heisenberg diamond}}
\centerline {\LARGE{spin-$1/2$ cluster in evolution}}
\medskip
\centerline {A. R. Kuzmak}
\centerline {\small \it E-Mail: andrijkuzmak@gmail.com}
\medskip
\centerline {\small \it Department for Theoretical Physics, Ivan Franko National University of Lviv,}
\medskip
\centerline {\small \it 12 Drahomanov St., Lviv, UA-79005, Ukraine}

\date{\today}
\begin{abstract}

In the last two decades, magnetic, thermodynamic properties and bipartite thermal entanglement in diamond spin clusters and chains have been studied. Such spin structures are presented in various compounds.
The ions of ${\rm Cu^{2+}}$ in the natural mineral azurite are arranged in a diamond spin chain. There are no studies of the entanglement behaviour during the quantum evolution of such systems. Herein,
we consider the evolution of entanglement in the diamond spin-1/2 cluster. This cluster consists of two central spins described by the anisotropic Heisenberg model, which interact with two side spins via Ising interaction.
The influence of the interaction coupling with side spins on the entanglement of central spins is investigated. It is shown that choosing the value of this coupling allows us to control the behaviour of entanglement
between central spins. As a result, we find conditions for achieving the maximal values of entanglement. In addition, the entanglement behaviour between the side spins, central and side spins,
and between a certain spin and the rest of the system is studied. In these cases, the conditions for achieving maximal entanglement are also obtained.

\end{abstract}

\section{Introduction}

Entanglement is the phenomenon inherent in a quantum system. Due to correlations between quantum particles \cite{einstein1935} it has no analogs in classical physics.
Aspect et al. experimentally tested Bell's inequality \cite{bell1964} solved the EPR paradox \cite{einstein1935} and thereby proved the existence of entanglement between quantum particles \cite{aspect982}.
Entanglement is an integral component that plays a central role in the implementation of quantum information schemes and devices.
The presence of entanglement in a quantum system allows us to realize quantum cryptography \cite{Ekert1991}, super-dense coding \cite{Bennett1992}, teleportation \cite{TELEPORT,Zeilinger1997},
quantum calculations \cite{cerf1998,pittman2001,gasparoni2004}, optimization of quantum calculations \cite{Giovannetti20031,Giovannetti20032,Batle2005,Borras2006}, etc. All these schemes require
the preparation of entanglement states on the physical systems. There are different quantum systems that are used for this purpose: polarized photons \cite{ASPECT,Zeilinger1997,gasparoni2004},
nuclear and electronic spins of atoms \cite{quantcomp,qdots1,phosphorus3,phosphorus1,kuzmak2020}, superconducting qubits \cite{supcond1,supcond2,supcond3,supcond4}, etc.
The behaviour of entanglement in such systems is important to investigate.

Herein we study the evolution of entanglement in the Ising-Heisenberg spin-$1/2$ diamond cluster. It was observed that the spins in compounds ${\rm Ca_3Cu_3(PO_4)_4}$, ${\rm Sr_3Cu_3(PO_4)_4}$ \cite{drillon1988,drillon1993}
${\rm Bi_4Cu_3V_2O_{14}}$ \cite{sakurai2002}, the natural mineral azurite (${\rm Cu_3(CO_3)_2(OH)_2}$) \cite{kikuchi2005} are arranged in diamond chains. For instance, the ${\rm Cu^{2+}}$ ions in the natural mineral azurite
form a spin-$1/2$ diamond chain. The different properties of the spin diamond model are studied in a wide range of papers. The magnetic properties such as magnetization and magnetic susceptibility
\cite{kikuchi2005,ananikian2012,Ishii2000,honecker2001,gu2007} and thermodynamic behaviour \cite{valverde2008,canova2006,gu2007,carvalho2019} of such models are well studied both theoretically and experimentally.

In the last twenty years, on the same level as magnetism and thermodynamics, the bipartite thermal entanglement in diamond spin clusters and chains has been actively studied
\cite{ananikian2012,bose2005,tribedi2006,ananikian2006,chakhmakhchyan2012,rojas2012,rojas2014,torrico2016,rojas2017,Zheng2018,Cavalho2019,Ghannadan2022}. One of the first works where the thermal entanglement in the spin
clusters was studied are the papers \cite{bose2005,tribedi2006}. In these systems, Bose and Tribedi calculated bipartite and multipartite entanglements as a function of temperature and signature of quantum phase transition,
in terms of the entanglement ratio. The thermal entanglement in the Ising-Heisenberg diamond chain was studied in papers \cite{ananikian2006,rojas2012,rojas2014}.
For the first time thermal entanglement of a spin-$1/2$ Ising-Heisenberg symmetrical diamond chain was studied by Nerses Ananikian et. all in paper \cite{ananikian2006}.
Thermal entanglement was calculated between spins in Heisenberg dimers separately. The authors studied in a wide range of coupling constant values the entanglement properties of the grodund state of the system.
They showed that for a dominant Heisenberg-type interaction the system's ground state is maximally entangled, but on increasing the temperature the pure quantum correlations disappear.
Onofre Rojas et. all calculated the concurrence as a function of temperature and external magnetic field of the Heisenberg dimer, which interacts with two nodal Ising spins.
They obtained results for two types of Heisenberg interaction, both $XXZ$-Heisenberg \cite{rojas2012} and $XYZ$-Heisenberg \cite{rojas2014} interactions, respectively. Thermal entanglement of the distorted diamond chain
model for azurite using pure quantum exchange interactions \cite{ananikian2012} and the Ising-XYZ distorted diamond chain with the second-neighbour interaction between nodal Ising spins \cite{rojas2017} was well studied.
The influence of the Dzyaloshinskii-Moriya interaction and impurities on the thermal entanglement in the spin-1/2 Ising-Heisenberg diamond chain was investigated in papers \cite{Zheng2018} and \cite{Cavalho2019}, respectively.
Recently, in paper \cite{Ghannadan2022} the thermal bipartite entanglement of a quantum spin-1 Heisenberg diamond cluster in the presence of an external magnetic field was quantified through the negativity.
The authors calculated thermal entanglement for spin pairs from the short diagonal and from the side of the diamond spin cluster. The results were applied to the spin-1 diamond cluster formed by ${\rm Ni^{2+}}$ ions
in the ${\rm [Ni_4(\mu-CO_3)_2(aetpy)_8](ClO_4)}_4$ compound, where aetpy = 2-aminoethyl-pyridine \cite{escuer1998,hagiwara2006}.

In the previous papers, the entanglement of the diamond spin systems in thermodynamic equilibrium was studied. However, there are no studies of the entanglement behaviour during the quantum evolution of such systems.
Since the presence of entanglement in the system is an inherent factor for the implementation of different schemes in quantum information (for instance, see \cite{desurvire2009}),
exploring entanglement behaviour and obtaining the conditions which allow the system achieves maximal entanglement is important. Quite a lot of these schemes, such as the implementation of quantum gates
on quantum computers, are based on the evolution of quantum systems. The diamond spin-1/2 cluster can be used for implementation schemes of quantum information that require the presence of entanglement.
Therefore, we study the behaviour of bipartite entanglement in evolution between spins in the Ising-Heisenberg diamond spin-1/2 cluster. We consider the diamond spin-$1/2$ cluster, where two central spins
are described by the anisotropic Heisenberg model, and interact with two side spins via Ising interaction. The influence of the side spins on the behaviour of entanglement of two central spins is studied.
It is shown that side spins have a qualitative effect on the entanglement of central spins. Choosing the interaction couplings between side and central spins allows us to control the evolution of the entanglement
between central spins. This fact allows us to find the conditions for achieving maximal entanglement. We also show that the ratio between interaction coupling has an effect on the temporal periodicity of
entanglement between central spins. In addition, the evolution of entanglement between the side spins, the central and side spins,
the certain spin and the rest of the system is studied.

The paper is organized as follows. The evolution of the Ising-Heisenberg diamond spin-1/2 cluster is described in Sec.~\ref{sec_model}. In Sec.~\ref{sec_evolution} the dynamics of bipartite entanglement
between different parts of such a cluster is calculated. Namely, it is investigated the influence of side spins on the entanglement
of Heisenberg spins (Subsec.~\ref{subsab}), entanglement between the side spins (Subsec.~\ref{subsys12}), the central and side spins (Subsec.~\ref{entbetweentwosubsys}), and
the certain spin and the rest of the system (Subsec.~\ref{entaspin1withother}). Conclusions are presented in Sec.~\ref{conc}.

\section{Model of the diamond spin cluster \label{sec_model}}

\begin{figure}[!!h]
\centerline{\includegraphics[scale=0.55, angle=0.0, clip]{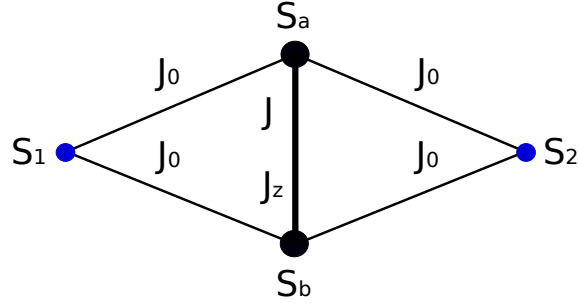}}
\caption{Model of the diamond spin-$1/2$ cluster. Two central spins $S_a$, $S_b$ are described by the anisotropic Heisenberg model, and interact with two side spins $S_1$, $S_2$ via Ising interaction.}
\label{model}
\end{figure}

We consider the system which consists of two central spin-$1/2$ $S_a$, $S_b$, described by anisotropic Heisenberg Hamiltonian, which interact with two side spin-$1/2$ $S_1$, $S_2$ via Ising interaction.
The system is placed in the magnetic field directed along the $z$-axis. The structure of this diamond spin cluster is shown in Fig.~\ref{model}.
Hamiltonian of the system has the form
\begin{eqnarray}
H=H_{ab}+H_{12}+H_{int},
\label{hamiltonian}
\end{eqnarray}
where
\begin{eqnarray}
&&H_{ab}=J\left(S_{a}^xS_{b}^x+S_{a}^yS_{b}^y\right)+J_zS_{a}^zS_{b}^z+h'\left(S_{a}^z+S_{b}^z\right),\label{hamab}\\
&&H_{12}=h\left(S_{1}^z+S_{2}^z\right),\label{ham12}\\
&&H_{int}=J_0\left(S_{a}^z+S_{b}^z\right)\left(S_{1}^z+S_{2}^z\right)\label{hamint}.
\end{eqnarray}
Here ${\bf S}_{\alpha}=1/2\left(S_{\alpha}^x{\bf i}+S_{\alpha}^y{\bf j}+S_{\alpha}^z{\bf k}\right)$ is the operator of $\alpha$-th spin ($\alpha=a,b,1,2$), $J$ and $J_z$ are the coupling constants between $a$ and $b$ spins,
$J_0$ is a coupling constant which defines the interaction between central $S_a$, $S_b$ and side $S_1$, $S_2$ spins, $h'$ and $h$ are the values which describe the interaction between spins and an external magnetic field.
We use the system of units, where the Planck constant is $\hbar =1$. This means that the energy is measured in units of frequency. Note that Hamiltonians (\ref{hamab}), (\ref{ham12}) and (\ref{hamint}) mutually
commute
\begin{eqnarray}
\left[H_{ab},H_{12}\right]=\left[H_{ab},H_{int}\right]=\left[H_{12},H_{int}\right]=0.\nonumber
\end{eqnarray}
This fact allows us to easily find the eigenvalues and eigenstates of the system. Thus Hamiltonian (\ref{hamiltonian}) has the following eigenvalues and eigenstates
\begin{align}
&\vert\psi_1\rangle = \vert\uparrow\uparrow\rangle_{{\tiny 12}}\vert\uparrow\uparrow\rangle_{\small{ab}}, && E_{1}=h+\frac{J_z}{4}+h'+J_0,\nonumber\\
&\vert\psi_2\rangle = \vert\uparrow\uparrow\rangle_{\small{12}}\frac{1}{\sqrt{2}}\left(\vert\uparrow\downarrow\rangle+\vert\downarrow\uparrow\rangle\right)_{ab}, && E_{2}=h+\frac{J}{2}-\frac{J_z}{4},\nonumber\\
&\vert\psi_3\rangle = \vert\uparrow\uparrow\rangle_{\small{12}}\frac{1}{\sqrt{2}}\left(\vert\uparrow\downarrow\rangle-\vert\downarrow\uparrow\rangle\right)_{ab}, && E_{3}=h-\frac{J}{2}-\frac{J_z}{4},\nonumber\\
&\vert\psi_4\rangle = \vert\uparrow\uparrow\rangle_{{\tiny 12}}\vert\downarrow\downarrow\rangle_{\small{ab}}, && E_{4}=h+\frac{J_z}{4}-h'-J_0,\nonumber\\
&\vert\psi_5\rangle = \vert\uparrow\downarrow\rangle_{{\tiny 12}}\vert\uparrow\uparrow\rangle_{\small{ab}}, && E_{5}=\frac{J_z}{4}+h',\nonumber\\
&\vert\psi_6\rangle = \vert\uparrow\downarrow\rangle_{\small{12}}\frac{1}{\sqrt{2}}\left(\vert\uparrow\downarrow\rangle+\vert\downarrow\uparrow\rangle\right)_{ab}, && E_{6}=\frac{J}{2}-\frac{J_z}{4},\nonumber\\
&\vert\psi_7\rangle = \vert\uparrow\downarrow\rangle_{\small{12}}\frac{1}{\sqrt{2}}\left(\vert\uparrow\downarrow\rangle-\vert\downarrow\uparrow\rangle\right)_{ab}, && E_{7}=-\frac{J}{2}-\frac{J_z}{4},\nonumber\\
&\vert\psi_8\rangle = \vert\uparrow\downarrow\rangle_{{\tiny 12}}\vert\downarrow\downarrow\rangle_{\small{ab}}, && E_{8}=\frac{J_z}{4}-h',\nonumber\\
&\vert\psi_9\rangle = \vert\downarrow\uparrow\rangle_{{\tiny 12}}\vert\uparrow\uparrow\rangle_{\small{ab}}, && E_{9}=\frac{J_z}{4}+h',\nonumber\\
&\vert\psi_{10}\rangle = \vert\downarrow\uparrow\rangle_{\small{12}}\frac{1}{\sqrt{2}}\left(\vert\uparrow\downarrow\rangle+\vert\downarrow\uparrow\rangle\right)_{ab}, && E_{10}=\frac{J}{2}-\frac{J_z}{4},\nonumber\\
&\vert\psi_{11}\rangle = \vert\downarrow\uparrow\rangle_{\small{12}}\frac{1}{\sqrt{2}}\left(\vert\uparrow\downarrow\rangle-\vert\downarrow\uparrow\rangle\right)_{ab}, && E_{11}=-\frac{J}{2}-\frac{J_z}{4},\nonumber\\
&\vert\psi_{12}\rangle = \vert\downarrow\uparrow\rangle_{{\tiny 12}}\vert\downarrow\downarrow\rangle_{\small{ab}}, && E_{12}=\frac{J_z}{4}-h',\nonumber\\
&\vert\psi_{13}\rangle = \vert\downarrow\downarrow\rangle_{{\tiny 12}}\vert\uparrow\uparrow\rangle_{\small{ab}}, && E_{13}=-h+\frac{J_z}{4}+h'-J_0,\nonumber\\
&\vert\psi_{14}\rangle = \vert\downarrow\downarrow\rangle_{\small{12}}\frac{1}{\sqrt{2}}\left(\vert\uparrow\downarrow\rangle+\vert\downarrow\uparrow\rangle\right)_{ab}, && E_{14}=-h+\frac{J}{2}-\frac{J_z}{4},\nonumber\\
&\vert\psi_{15}\rangle = \vert\downarrow\downarrow\rangle_{\small{12}}\frac{1}{\sqrt{2}}\left(\vert\uparrow\downarrow\rangle-\vert\downarrow\uparrow\rangle\right)_{ab}, && E_{15}=-h-\frac{J}{2}-\frac{J_z}{4},\nonumber\\
&\vert\psi_{16}\rangle = \vert\downarrow\downarrow\rangle_{{\tiny 12}}\vert\downarrow\downarrow\rangle_{\small{ab}}, && E_{16}=-h+\frac{J_z}{4}-h'+J_0.
\label{eigenvaleigenstate}
\end{align}
The state of subsystems is indicated by the subscripts. States of $S_1$, $S_2$ and $S_a$, $S_b$ spins are denoted by the subscripts $12$ and $ab$, respectively.

Evolution of a system defined by Hamiltonian (\ref{hamiltonian}) having started from the initial state $\vert\psi_I\rangle$ can be expressed as follows
\begin{eqnarray}
\vert\psi(t)\rangle=e^{-iHt}\vert\psi_I\rangle=e^{-iHt}\sum_n C_n\vert\psi_n\rangle=\sum_n C_ne^{-iE_nt}\vert\psi_n\rangle,
\label{evolution}
\end{eqnarray}
where $\vert\psi_n\rangle$ and $E_n$ is a predefined set of eigenstates and eigenvalues given by expression (\ref{eigenvaleigenstate}), $C_n$ are the complex parameters which determine the initial state.
We decompose the initial state in terms of eigenstates (\ref{eigenvaleigenstate}). Let us study the entanglement of states achieved in this system.

\section{Entanglement in the diamond spin cluster \label{sec_evolution}}

In this section, we calculate the value of entanglement between different parts of the spin cluster described by Hamiltonian (\ref{hamiltonian}).
To calculate the entanglement of state $\rho$, we use the Wootters definition of concurrence \cite{wootters1998}
\begin{eqnarray}
C(\rho)=\max\{0,\omega_1-\omega_2-\omega_3-\omega_4\},
\label{wootters}
\end{eqnarray}
where $\omega_i$ are the eigenvalues, in decreasing order, of the Hermitian matrix $R=\sqrt{\sqrt{\rho}\tilde{\rho}\sqrt{\rho}}$. Here,
$\tilde{\rho}= \sigma^y\otimes\sigma^y \rho^*\sigma^y\otimes\sigma^y$. Note that $\omega_i$ are real and positive numbers.
For calculations, it is convenient to use the eigenvalues of the non-Hermitian matrix $\rho\tilde{\rho}$ which have the form $\omega_i^2$. In the case of a pure two-qubit state
\begin{eqnarray}
\vert\psi\rangle =a\vert\uparrow\uparrow\rangle+b\vert\uparrow\downarrow\rangle+c\vert\downarrow\uparrow\rangle+d\vert\downarrow\downarrow\rangle
\label{purestate}
\end{eqnarray}
definition (\ref{wootters}) takes the form \cite{wootters1997}
\begin{eqnarray}
C(\vert\psi\rangle)=2\vert ad-bc\vert,
\label{wootterspure}
\end{eqnarray}
where $a$, $b$, $c$ and $d$ complex parameters which satisfy normalization condition $\vert a\vert^2+\vert b\vert^2+\vert c\vert^2+\vert d\vert^2=1$.

The most general definition of bipartite entanglement of pure state $\rho=\vert\psi\rangle\langle\psi\vert$ is von Neumann entropy of the subsystems \cite{wootters19960,popescu1997}. Suppose that a quantum system consists
of two subsystems $A$ and $B$. The value of entanglement is defined with respect to these subsystems as follows
\begin{eqnarray}
E(\vert\psi\rangle)=-{\rm Tr}\left(\rho_A\log\rho_A\right)=-{\rm Tr}\left(\rho_B\log\rho_B\right),
\label{entropyofentanglement}
\end{eqnarray}
where $\rho_{A(B)}$ is the partial trace of $\rho$ over the subsystem $B(A)$. In the case of mixed state $\rho$ consisting of the pure states $\vert\psi_i\rangle$, the bipartite measure of entanglement
can be obtained using entanglement of formation \cite{bennett1996}. It is defined by the average entanglement of the pure states $\vert\psi_i\rangle$ of the decomposition minimized over all decompositions of mixed states $\rho$
with probabilities $p_i$ in the form
\begin{eqnarray}
E(\rho)=\min\sum_ip_i E(\vert\psi_i\rangle).
\label{entangledofformation}
\end{eqnarray}
For a two-qubit system, the entanglement of formation is a monotonic function of concurrence (\ref{wootters})
\begin{eqnarray}
&&E(\rho)=-\frac{1+\sqrt{1-C(\rho)^2}}{2}\log_2\left(\frac{1+\sqrt{1-C(\rho)^2}}{2}\right)\nonumber\\
&&-\frac{1-\sqrt{1-C(\rho)^2}}{2}\log_2\left(\frac{1-\sqrt{1-C(\rho)^2}}{2}\right).
\label{entofformationconcu}
\end{eqnarray}

\subsection{Entanglement between the $S_a$ and $S_b$ spins \label{subsab}}

In this section we consider the evolution of the system having started from the initial state when all spins are projected on the positive direction of the $x$-axis. This state can be expressed as follows
\begin{eqnarray}
&&\vert\psi_I\rangle=\frac{1}{4}\left(\vert\uparrow\uparrow\rangle_{12}+\vert\uparrow\downarrow\rangle_{12}+\vert\downarrow\uparrow\rangle_{12}+\vert\downarrow\downarrow\rangle_{12}\right)\nonumber\\
&&\left(\vert\uparrow\uparrow\rangle_{ab}+\vert\uparrow\downarrow\rangle_{ab}+\vert\downarrow\uparrow\rangle_{ab}+\vert\downarrow\downarrow\rangle_{ab}\right).
\label{initstate12ab}
\end{eqnarray}
It can be expressed by eigenstates (\ref{eigenvaleigenstate}) as follows
\begin{eqnarray}
&&\vert\psi_I\rangle=\frac{1}{4}\left(\vert\psi_1\rangle+\sqrt{2}\vert\psi_2\rangle+\vert\psi_4\rangle+\vert\psi_5\rangle+\sqrt{2}\vert\psi_6\rangle+\vert\psi_8\rangle\right.\nonumber\\
&&\left.+ \vert\psi_9\rangle+\sqrt{2}\vert\psi_{10}\rangle+\vert\psi_{12}\rangle+\vert\psi_{13}\rangle+\sqrt{2}\vert\psi_{14}\rangle+\vert\psi_{16}\rangle\right).
\label{initstate12ab2}
\end{eqnarray}
Using expression (\ref{evolution}) the evolution of the system takes the form
\begin{eqnarray}
&&\vert\psi(t)\rangle=e^{-iHt}\vert\psi_I\rangle=\frac{1}{4}\left(e^{-iE_1t}\vert\psi_1\rangle+\sqrt{2}e^{-iE_2t}\vert\psi_2\rangle + e^{-iE_4t}\vert\psi_4\rangle \right.\nonumber\\
&&\left.+e^{-iE_5t}\vert\psi_5\rangle+\sqrt{2}e^{-iE_6t}\vert\psi_6\rangle+e^{-iE_8t}\vert\psi_8\rangle+ e^{-iE_9t}\vert\psi_9\rangle+\sqrt{2}e^{-iE_{10}t}\vert\psi_{10}\rangle\right.\nonumber\\
&&\left.+e^{-iE_{12}t}\vert\psi_{12}\rangle+e^{-iE_{13}t}\vert\psi_{13}\rangle+\sqrt{2}e^{-iE_{14}t}\vert\psi_{14}\rangle+e^{-iE_{16}t}\vert\psi_{16}\rangle\right),
\label{evolution12ab}
\end{eqnarray}
where eigenvalues $E_i$ are defined by equation (\ref{eigenvaleigenstate}). To study the evolution of subsystem determined by spins $S_a$ and $S_b$ we represent this state in the form
\begin{eqnarray}
\vert\psi(t)\rangle=\frac{1}{2}\left(\vert\xi_1\rangle_{ab}\vert\uparrow\uparrow\rangle_{12}+ \vert\xi_2\rangle_{ab}\left(\vert\uparrow\downarrow\rangle_{12}+\vert\downarrow\uparrow\rangle_{12}\right) + \vert\xi_3\rangle_{ab}\vert\downarrow\downarrow\rangle_{12}\right),
\label{evolution12ab2}
\end{eqnarray}
where we introduce the following notation for the state of spins $S_a$ and $S_b$
{\small
\begin{eqnarray}
&&\vert\xi_1\rangle_{ab}=\frac{1}{2}\left[e^{-i(\frac{J_z}{4}+J_0+h+h')t}\vert\uparrow\uparrow\rangle_{ab}+e^{-i(\frac{J}{2}-\frac{J_z}{4}+h)t}(\vert\uparrow\downarrow\rangle_{ab}+\vert\downarrow\uparrow\rangle_{ab})+e^{-i(\frac{J_z}{4}-J_0+h-h')t}\vert\downarrow\downarrow\rangle_{ab}     \right],\nonumber\\
&&\vert\xi_2\rangle_{ab}=\frac{1}{2}\left[e^{-i(\frac{J_z}{4}+h')t}\vert\uparrow\uparrow\rangle_{ab}+e^{-i(\frac{J}{2}-\frac{J_z}{4})t}(\vert\uparrow\downarrow\rangle_{ab}+\vert\downarrow\uparrow\rangle_{ab})+e^{-i(\frac{J_z}{4}-h')t}\vert\downarrow\downarrow\rangle_{ab}     \right],\nonumber\\
&&\vert\xi_3\rangle_{ab}=\frac{1}{2}\left[e^{-i(\frac{J_z}{4}-J_0-h+h')t}\vert\uparrow\uparrow\rangle_{ab}+e^{-i(\frac{J}{2}-\frac{J_z}{4}-h)t}(\vert\uparrow\downarrow\rangle_{ab}+\vert\downarrow\uparrow\rangle_{ab})+e^{-i(\frac{J_z}{4}+J_0-h-h')t}\vert\downarrow\downarrow\rangle_{ab}     \right].\nonumber\\
\label{purestateofabspins}
\end{eqnarray}}
\begin{figure}[!!h]
\subfloat[]{\label{ent_abspins_J0_0.3J}}\includegraphics[scale=0.53, angle=0.0, clip]{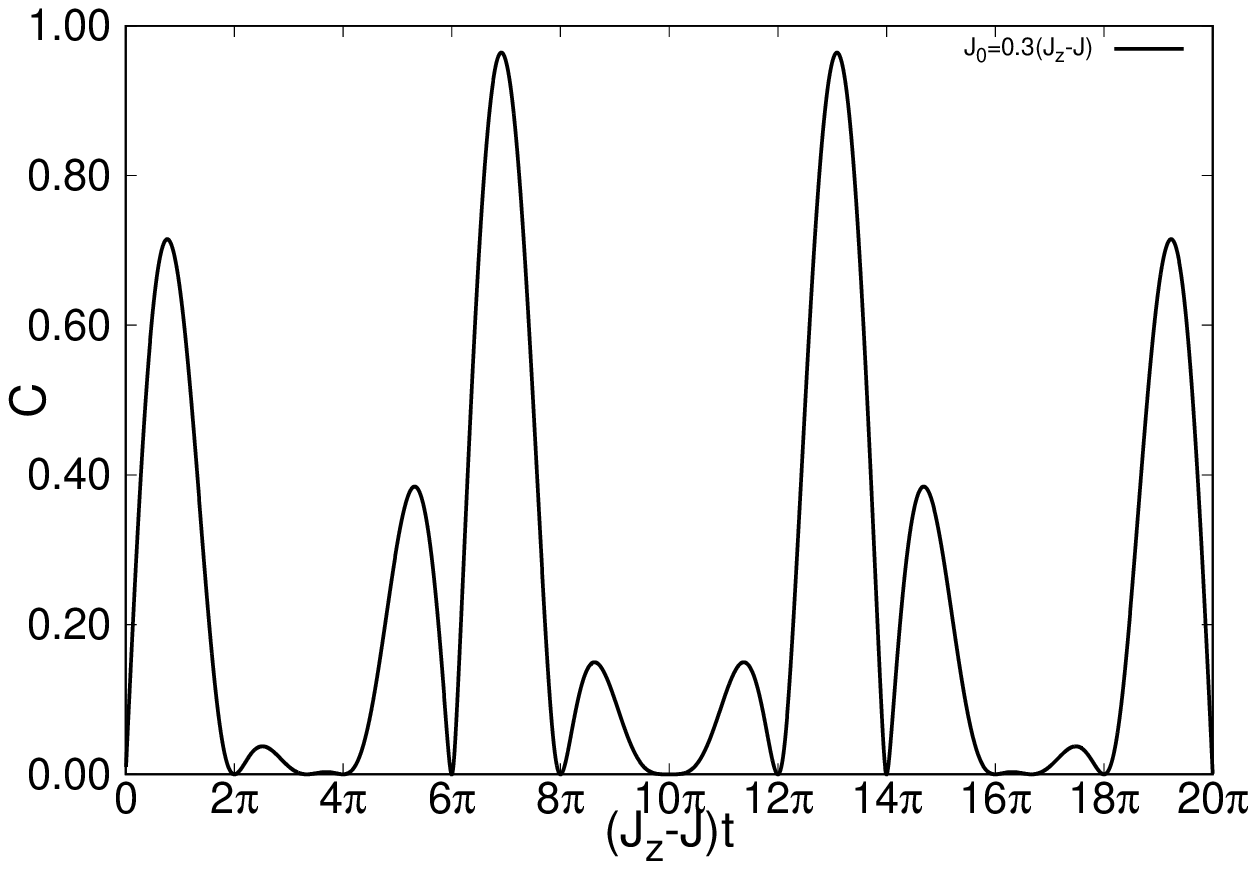}\includegraphics[scale=0.53, angle=0.0, clip]{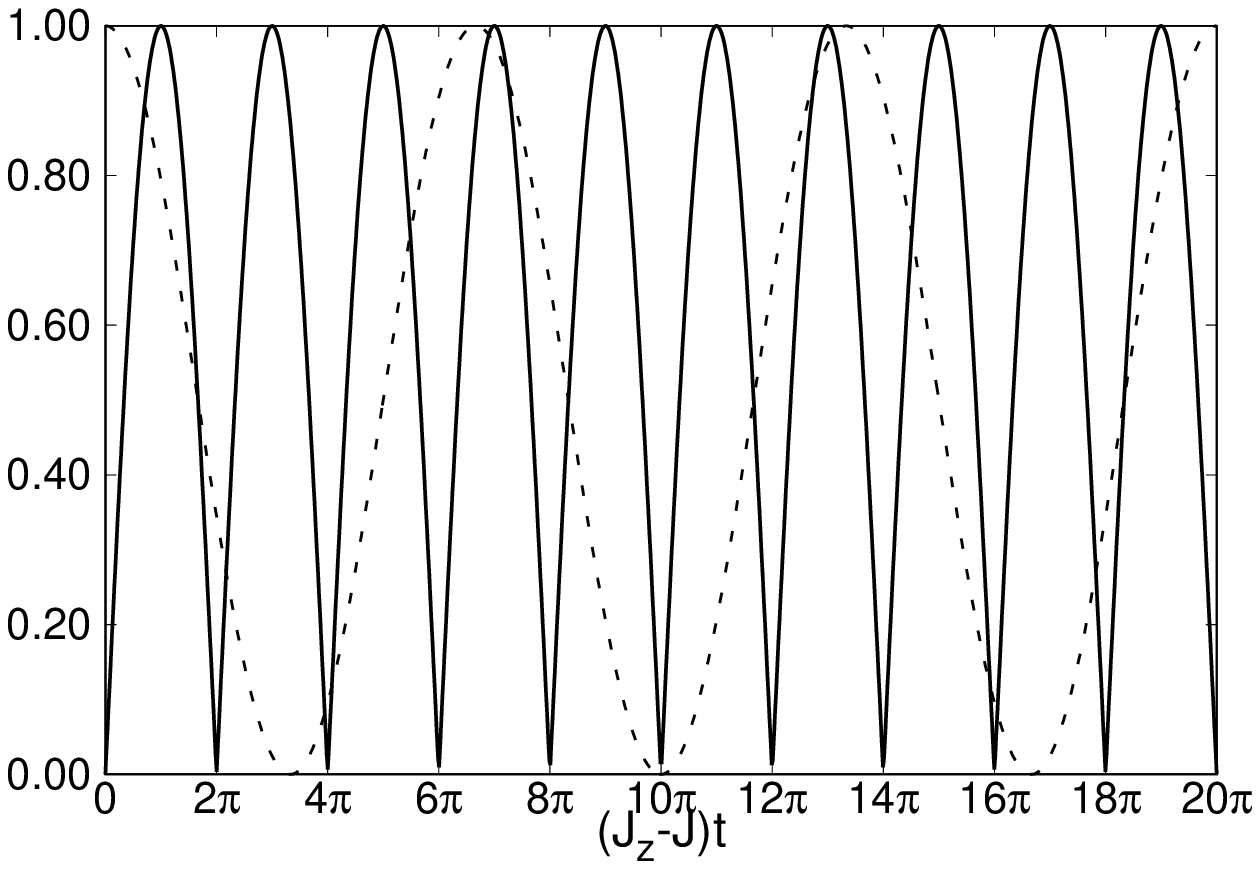}
\subfloat[]{\label{ent_abspins_J0_10.0J}}\includegraphics[scale=0.53, angle=0.0, clip]{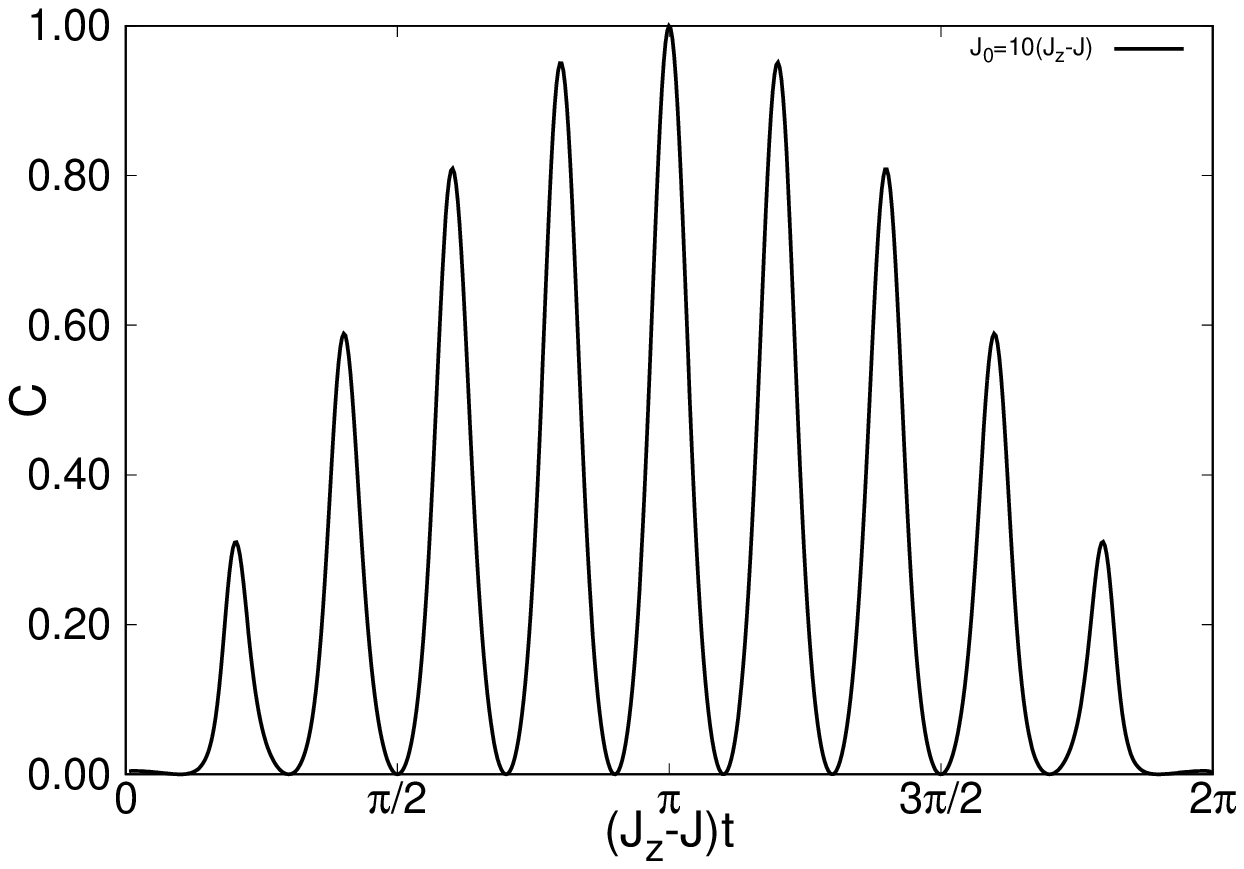}\includegraphics[scale=0.53, angle=0.0, clip]{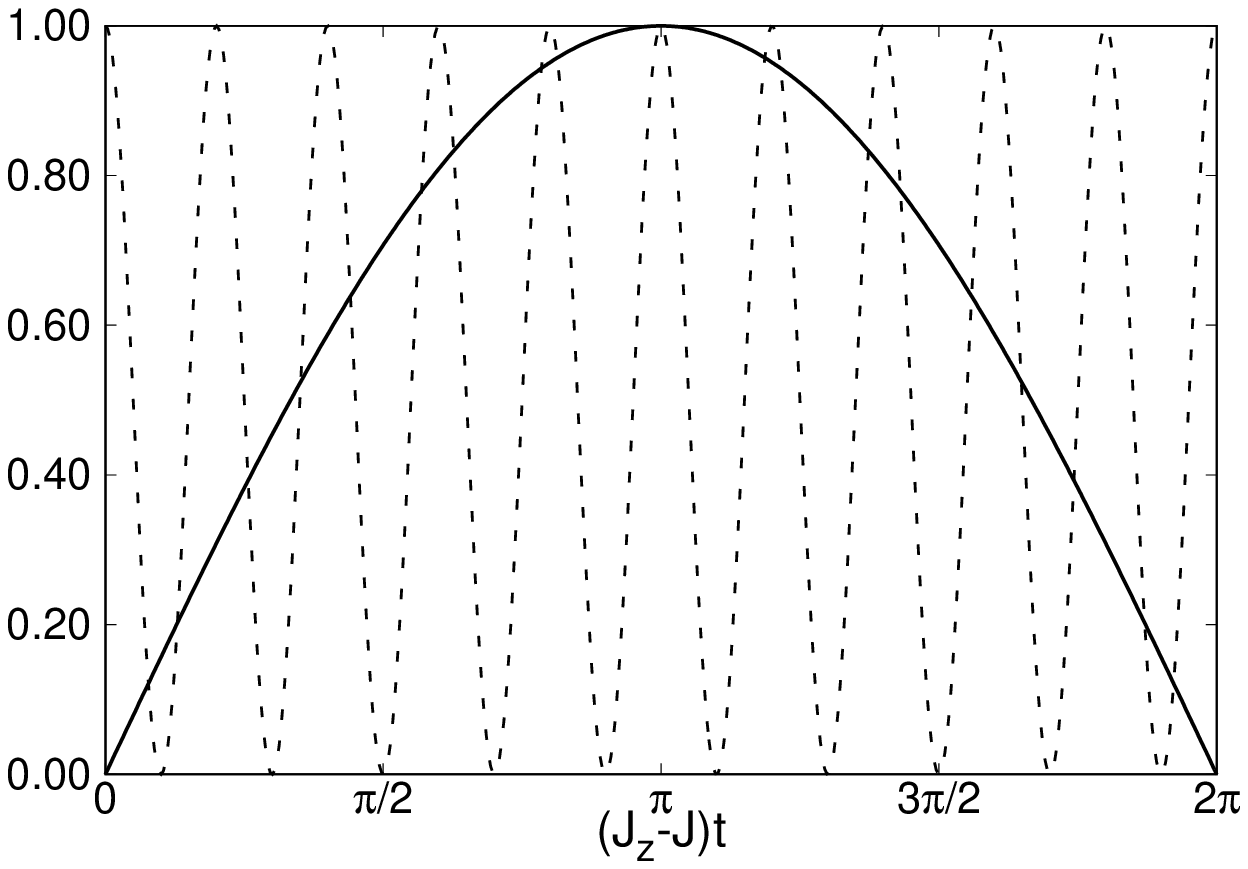}
\caption{In the left column, the time dependencies of concurrence for $J_0=0.3(J_z-J)$ (a) and $J_0=10(J_z-J)$ (b) are shown. The behaviour of these dependencies is defined by the parameter $J_0$
as a function of time $\cos^2\frac{J_0t}{2}$. Thus at the minimum of this function, the value of entanglement between spins decreases, and at the maximum, it increases. In the right column,
the corresponding effect for $J_0=0.3(J_z-J)$ (a) and $J_0=10(J_z-J)$ (b) is presented. The solid line depicts the concurrence between spins for the case of $J_0=0$, and the dashed line
depicts behaviour of function $\cos^2\frac{J_0t}{2}$.}
\label{ent_abspins_J0}
\end{figure}
\begin{figure}[!!h]
\subfloat[]{\label{ent_abspins_J0_0.01J}}\includegraphics[scale=0.54, angle=0.0, clip]{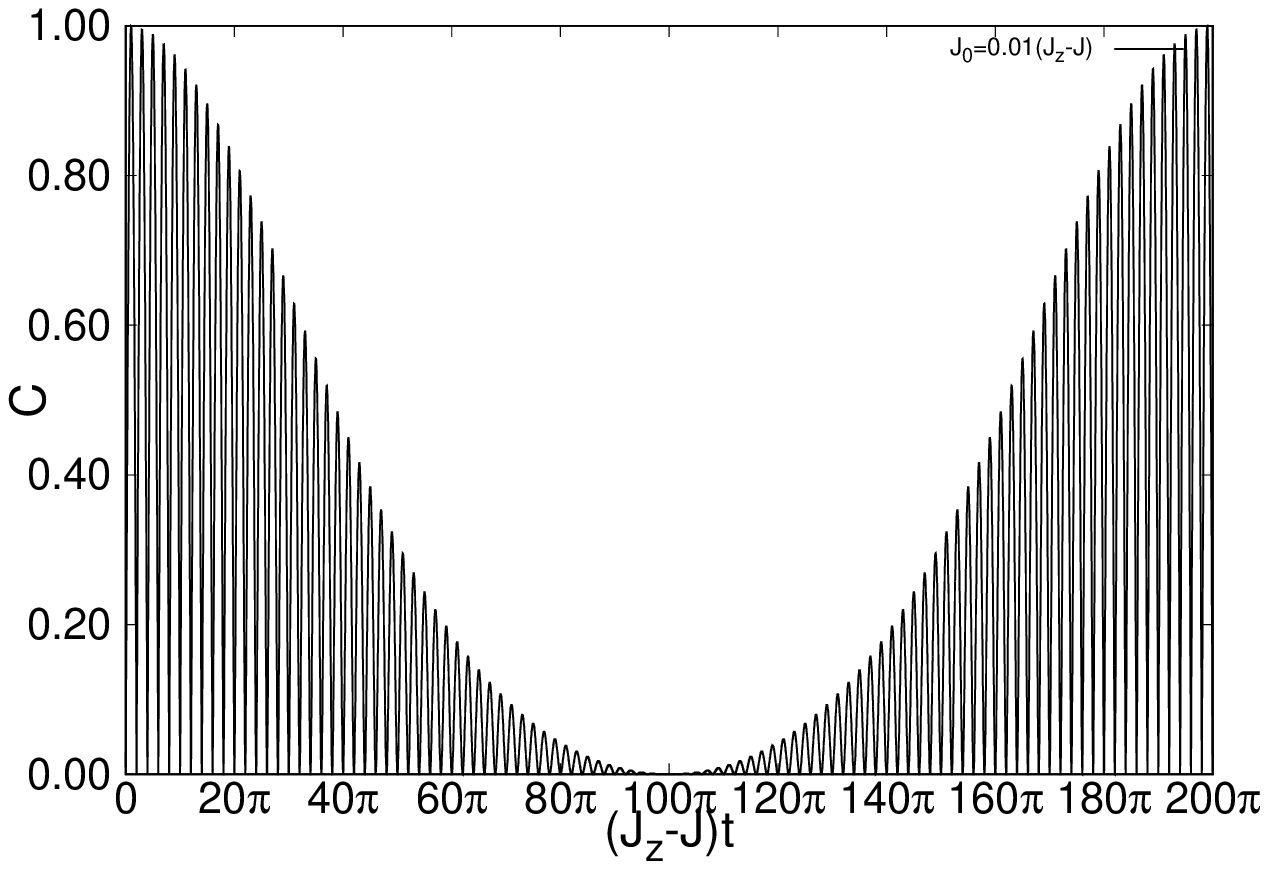}
\subfloat[]{\label{ent_abspins_J0_0.1J}}\includegraphics[scale=0.54, angle=0.0, clip]{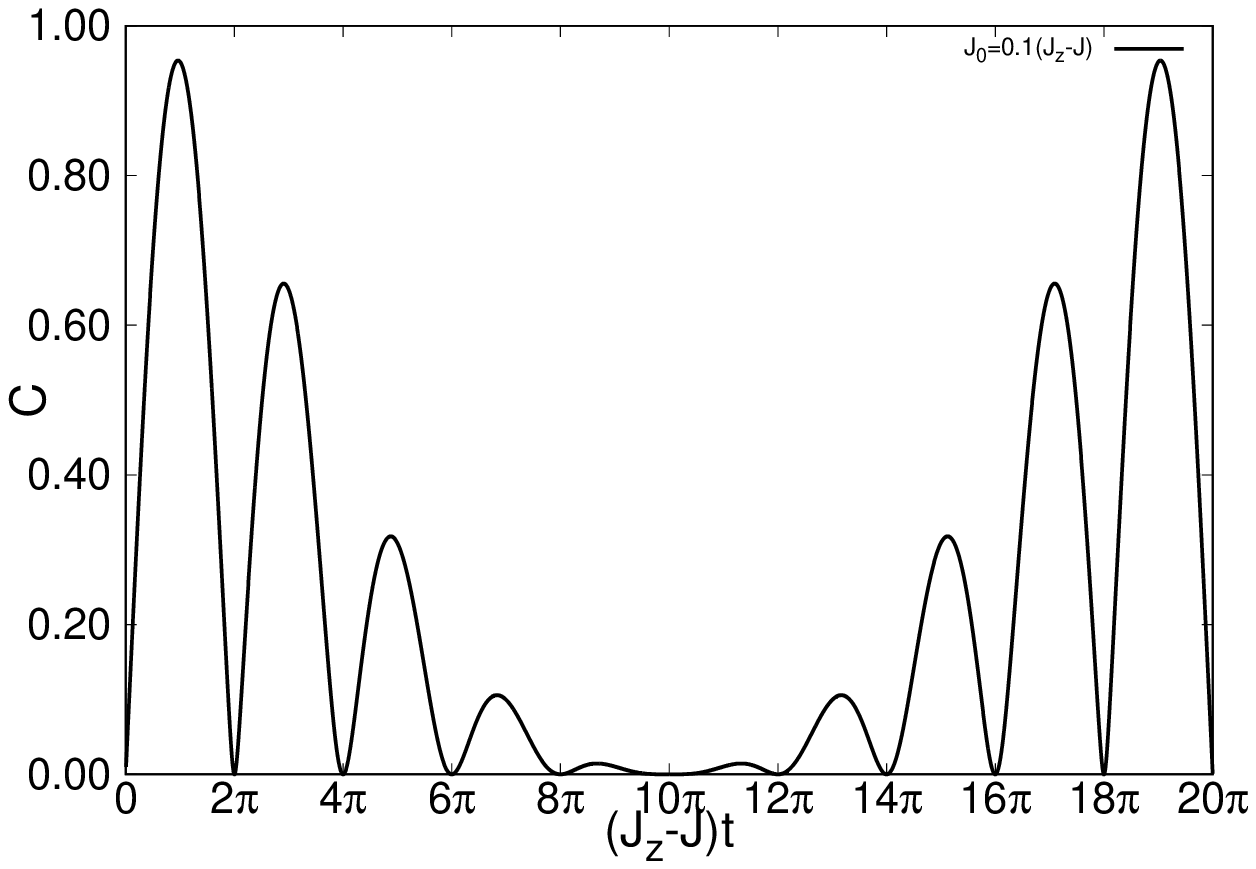}
\subfloat[]{\label{ent_abspins_J0_0.9J}}\includegraphics[scale=0.54, angle=0.0, clip]{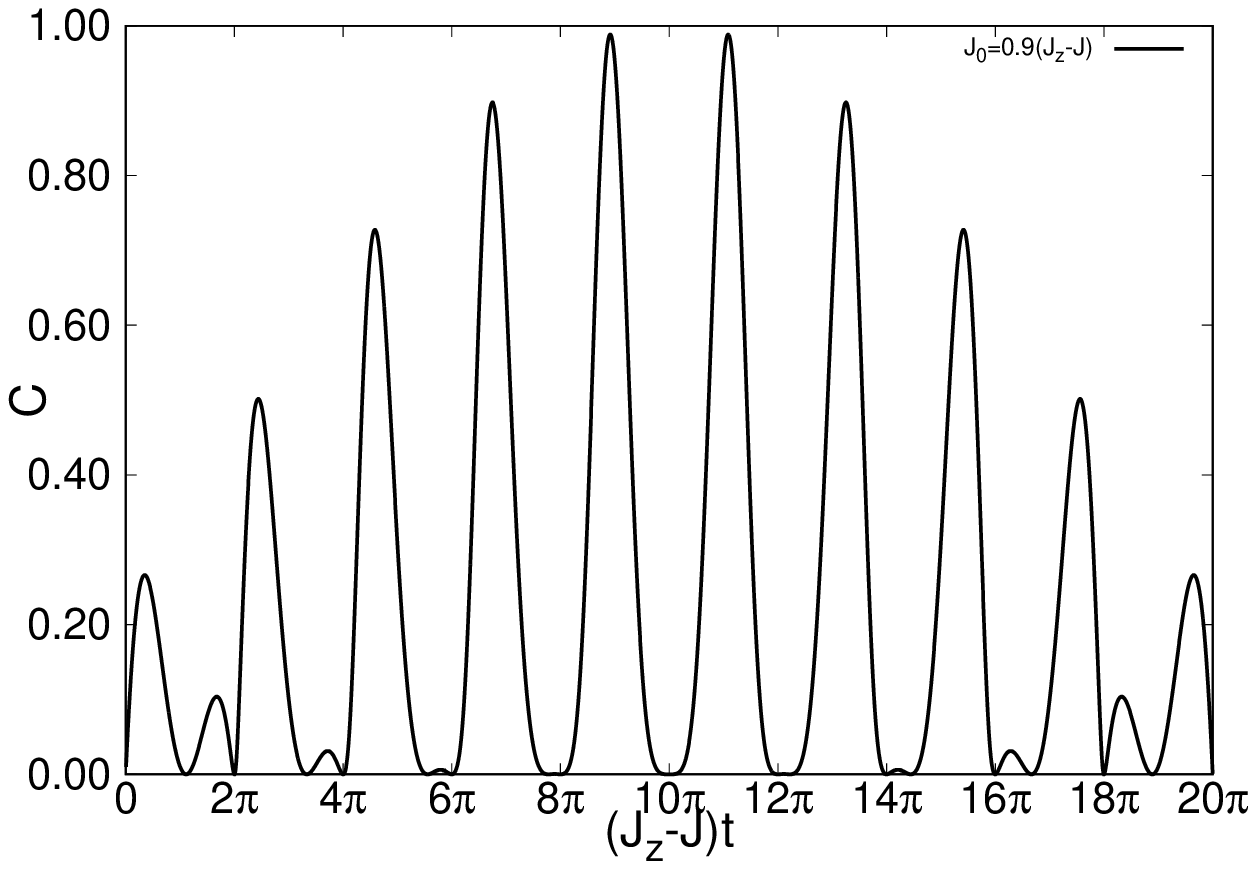}
\subfloat[]{\label{ent_abspins_J0_0.99J}}\includegraphics[scale=0.54, angle=0.0, clip]{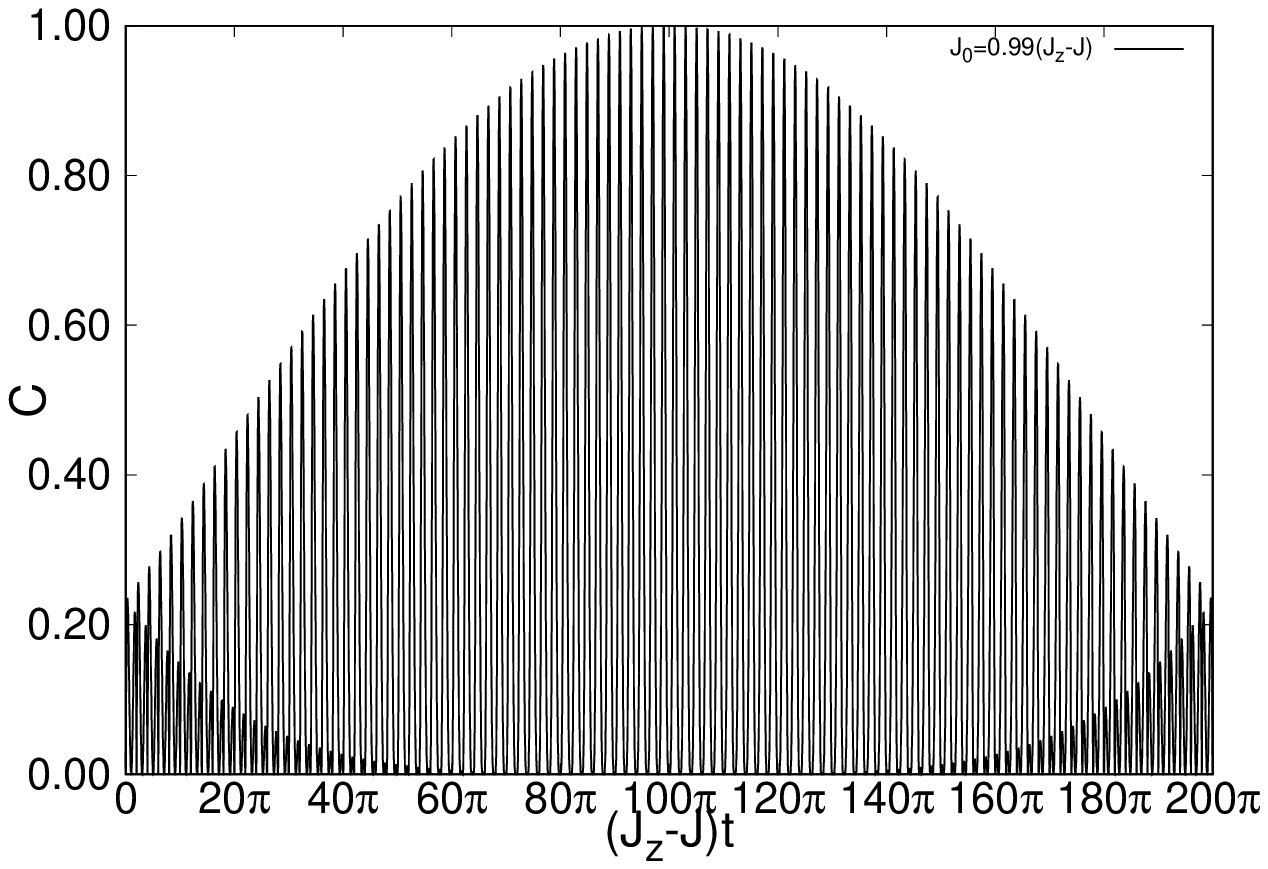}
\caption{The time dependencies of concurrence between spins $S_a$ and $S_b$ depend on the ratio between parameters $J_0$ and $J_z-J$. As we can see that the fractional nature of the ratio affects
the periodicity of entanglement behaviour. In the cases of $J_0=0.01(J_z-J)$ (a), $J_0=0.99(J_z-J)$ (d) the temporal periodicity of entanglement is $200\pi$, and in the cases of $J_0=0.1(J_z-J)$ (b)
$J_0=0.9(J_z-J)$ (c) the temporal periodicity of entanglement is $20\pi$.}
\label{ent_abspins_J02}
\end{figure}
\begin{figure}[!!h]
\subfloat[]{\label{ent_abspins_J0_1.0J}}\includegraphics[scale=0.54, angle=0.0, clip]{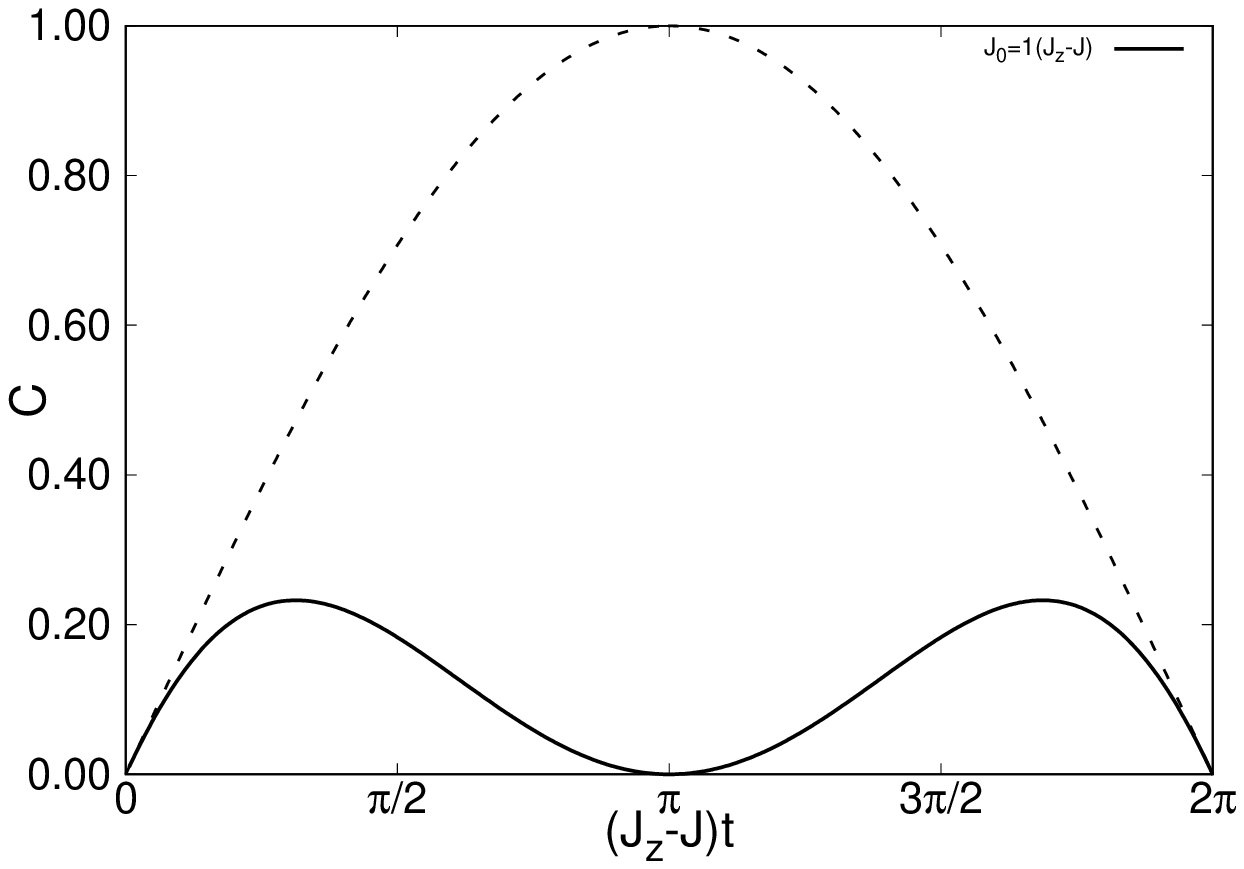}
\subfloat[]{\label{ent_abspins_J0_2.0J}}\includegraphics[scale=0.54, angle=0.0, clip]{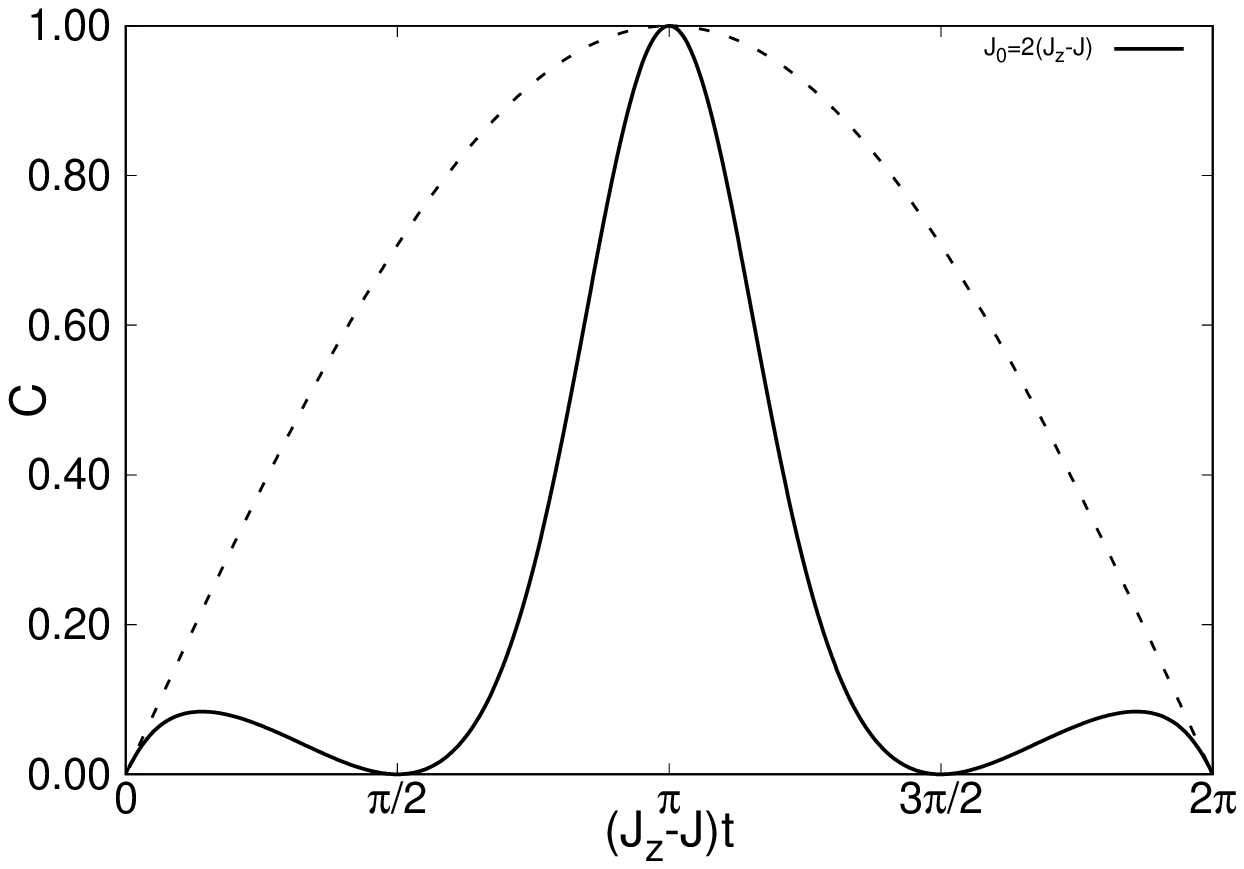}
\subfloat[]{\label{ent_abspins_J0_5.0J}}\includegraphics[scale=0.54, angle=0.0, clip]{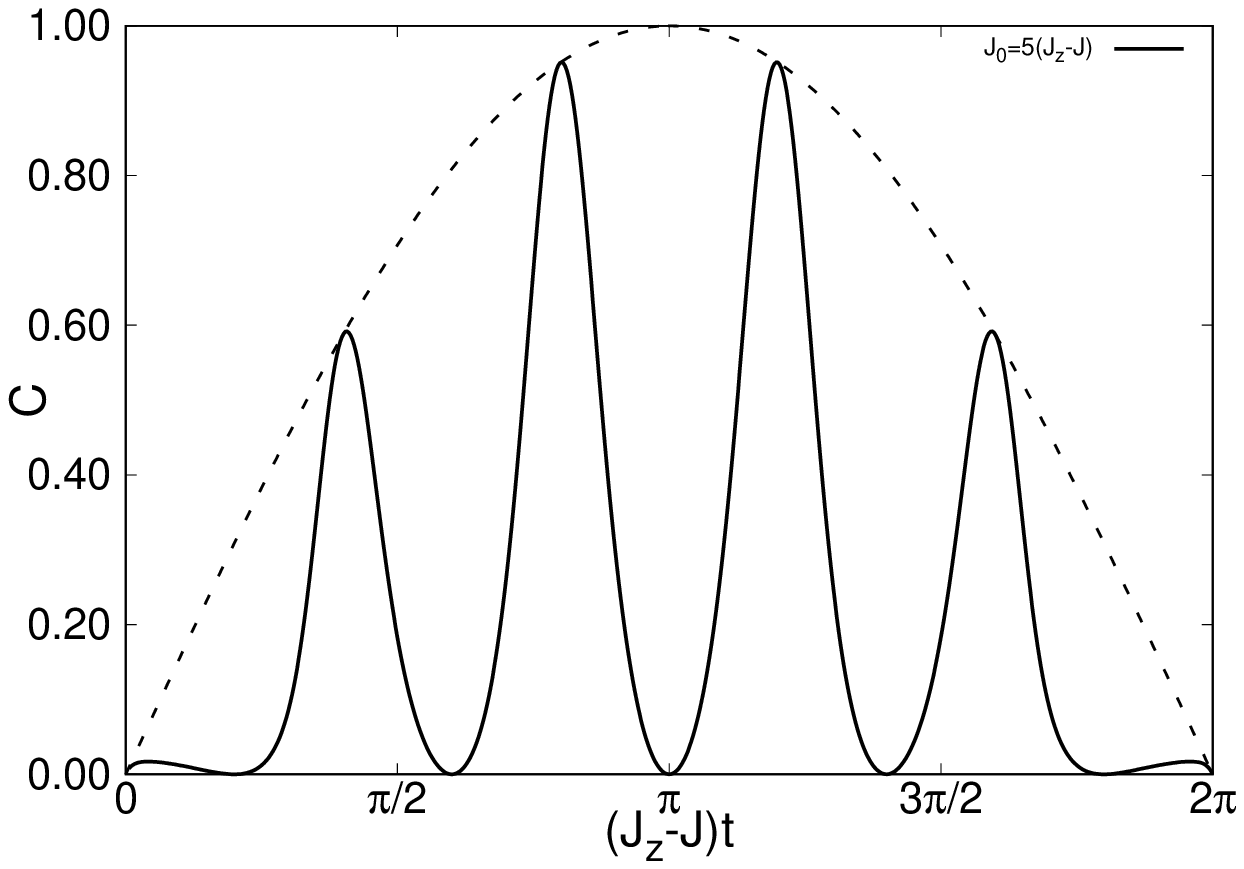}
\subfloat[]{\label{ent_abspins_J0_20.0J}}\includegraphics[scale=0.54, angle=0.0, clip]{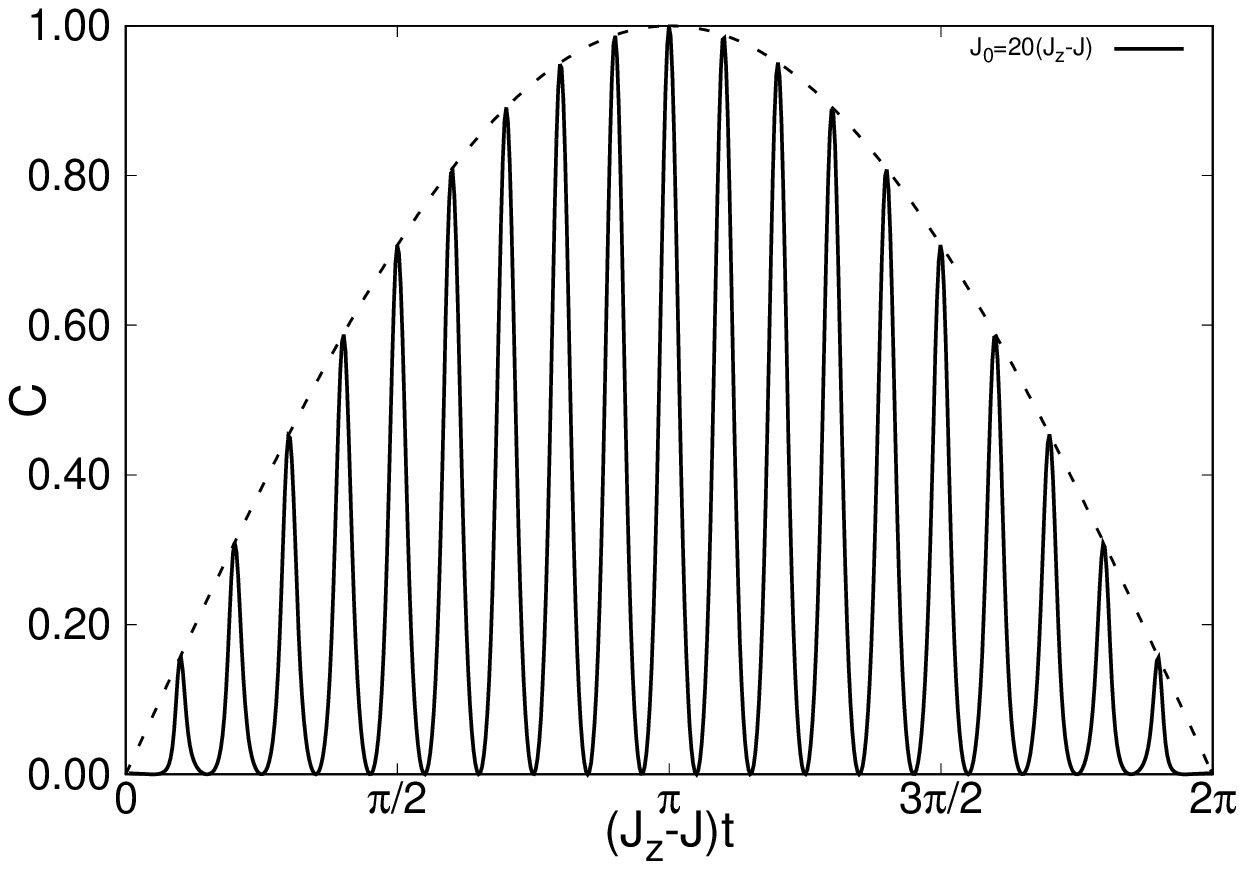}
\caption{The time dependencies of concurrence between spins $S_a$ and $S_b$ depend on the integer ratio between parameters $J_0$ and $J_z-J$. The dashed line depicts behaviour of concurrence in the case of $J_0=0$.}
\label{ent_abspins_J03}
\end{figure}
Here we use the explicit form of eigenvalues $E_i$. Let us study the entanglement between $S_a$ and $S_b$ spins. For this purpose, we average the density matrix of the system $\rho(t)=\vert\psi(t)\rangle\langle\psi(t)\vert$
over the states of the $S_1$ and $S_2$ spins
\begin{eqnarray}
\rho(t)_{ab}={\rm Tr}_{12}\rho(t)=\frac{1}{4}\left(\vert\xi_1\rangle_{ab}\langle\xi_1\vert_{ab}+2\vert\xi_2\rangle_{ab}\langle\xi_2\vert_{ab}+\vert\xi_3\rangle_{ab}\langle\xi_3\vert_{ab}\right).
\label{spinabdensmatrix}
\end{eqnarray}
As we can see the mixed state of $S_a$ and $S_b$ spins consists of the ensemble of pure states defined by formula (\ref{purestateofabspins}). The density matrix in the basis
$\vert\uparrow\uparrow\rangle$, $\vert\uparrow\downarrow\rangle$, $\vert\downarrow\uparrow\rangle$, $\vert\downarrow\downarrow\rangle$ reads
\begin{eqnarray}
\rho(t)_{ab}=\left( \begin{array}{ccccc}
\frac{1}{4} & \frac{1}{8}e^{-ih't}A(1+B) & \frac{1}{8}e^{-ih't}A(1+B) & \frac{1}{4}e^{-i2h't}B^2\\[9pt]
\frac{1}{8}e^{ih't}A^*(1+B) & \frac{1}{4} & \frac{1}{4} & \frac{1}{8}e^{-ih't}A^*(1+B) \\[9pt]
\frac{1}{8}e^{ih't}A^*(1+B) & \frac{1}{4} & \frac{1}{4} & \frac{1}{8}e^{-ih't}A^*(1+B) \\[9pt]
\frac{1}{4}e^{i2h't}B^2 & \frac{1}{8}e^{ih't}A(1+B) & \frac{1}{8}e^{ih't}A(1+B) & \frac{1}{4}
\end{array}\right),
\label{appa11}
\end{eqnarray}
where we introduce the following notations $A=e^{-i(J_z/2-J/2)t}$, $B=\cos(J_0t)$. Using Wootters definition of measure of entanglement (\ref{wootters})
we calculate the value of entanglement between spins $S_a$ and $S_b$ in state (\ref{spinabdensmatrix}). We presented the details of these calculations in Appendix \ref{appa}. As we can see that entanglement
between spins $S_a$ and $S_b$ is defined by the difference between the interaction couplings $J_z-J$ and depends on the influence of coupling $J_0$. The greater the difference $J_z-J$, the faster the entanglement
grows in the system. Note that in the case of $J_z-J=0$ the value of entanglement is zero all the time. If $J_0=0$ the spins $S_a$ and $S_b$ evolve without the influence of spins $S_1$ and $S_2$.
Thus solving equation (\ref{appa14}) with $J_0=0$ we obtain the concurrence in form
\begin{eqnarray}
C=\left\vert\sin\frac{J_z-J}{2}t\right\vert.
\label{concurrencepureabstate}
\end{eqnarray}
From the analysis of density matrix (\ref{appa11}) follows that spins $S_1$ and $S_2$ have a qualitative effect on the entanglement of spins $S_a$ and $S_b$ by the parameter $J_0$ as follows: $\frac{1}{2}(1+B)=\cos^2\frac{J_0t}{2}$.
By choosing a ratio between parameter $J_0$ and $J_z-J$, we can control the behaviour of the entanglement between spins $S_a$ and $S_b$. Thus at the minimum of function $\cos^2\frac{J_0t}{2}$, the value of entanglement between spins decreases,
and vice versa at the maximum, it increases. It is easy to determine the points at which the entanglement is zero. These points correspond to the vanishing of expressions (\ref{concurrencepureabstate})
and $\cos^2\frac{J_0t}{2}$. They read as follows
\begin{eqnarray}
(J_z-J)t_{min}=2n\pi,\quad J_0t_{min}=(2n+1)\pi,
\label{minpointofent}
\end{eqnarray}
where $n=0,1,2,\ldots$. Points with maximum entanglement values are determined from the equality of units of expressions (\ref{concurrencepureabstate}) and $\cos^2\frac{J_0t}{2}$.
Except for the points where at least one of those expressions vanishes, the points where entanglement has a local maximum determine as follows
\begin{eqnarray}
(J_z-J)t_{max}=(2n+1)\pi,\quad J_0t_{max}=2n\pi.
\label{maxpointofent}
\end{eqnarray}
In the cases where these points coincide with points defined by expression (\ref{minpointofent}), entanglement always vanishes.
In Fig.~\ref{ent_abspins_J0}, we quantitatively and qualitatively depicted this effect in the cases of $J_0=0.3(J_z-J)$ (Fig.~\ref{ent_abspins_J0_0.3J})
and $J_0=10(J_z-J)$ (Fig.~\ref{ent_abspins_J0_10.0J}). As we can see, depending on the ratio between parameters $J_0$ and $J_z-J$ we can control the value of entanglement during the evolution of the system.
This ratio also affects on periodicity of entanglement. In Fig.~\ref{ent_abspins_J02} we express how the fractional nature of this ratio affects the temporal periodicity of entanglement.
In the case of integer values of the ratios between parameters $J_0$ and $J_z-J$ we always have the periodicity of entanglement $2\pi$ (Fig.~\ref{ent_abspins_J03}).

\subsection{Entanglement of the $S_1$ and $S_2$ subsystem \label{subsys12}}

The state achieved during the evolution (\ref{evolution12ab}) we rewrite in the form
\begin{eqnarray}
\vert\psi(t)\rangle=\frac{1}{2}\left(\vert\phi_1\rangle_{12}\vert\uparrow\uparrow\rangle_{ab}+ \vert\phi_2\rangle_{12}\left(\vert\uparrow\downarrow\rangle_{ab}+\vert\downarrow\uparrow\rangle_{ab}\right) + \vert\phi_3\rangle_{12}\vert\downarrow\downarrow\rangle_{ab}\right),
\label{initstate12abevolut}
\end{eqnarray}
where we introduce the following notation
{\small
\begin{eqnarray}
&&\vert\phi_1\rangle_{12}=\frac{1}{2}\left[e^{-i(\frac{J_z}{4}+J_0+h+h')t}\vert\uparrow\uparrow\rangle_{12}+e^{-i(\frac{J_z}{4}+h')t}(\vert\uparrow\downarrow\rangle_{12}+\vert\downarrow\uparrow\rangle_{12})+e^{-i(\frac{J_z}{4}-J_0-h+h')t}\vert\downarrow\downarrow\rangle_{12}     \right],\nonumber\\
&&\vert\phi_2\rangle_{12}=\frac{1}{2}\left[e^{-i(\frac{J}{2}-\frac{J_z}{4}+h)t}\vert\uparrow\uparrow\rangle_{12}+e^{-i(\frac{J}{2}-\frac{J_z}{4})t}(\vert\uparrow\downarrow\rangle_{12}+\vert\downarrow\uparrow\rangle_{12})+e^{-i(\frac{J}{2}-\frac{J_z}{4}-h)t}\vert\downarrow\downarrow\rangle_{12}     \right],\nonumber\\
&&\vert\phi_3\rangle_{12}=\frac{1}{2}\left[e^{-i(\frac{J_z}{4}-J_0+h-h')t}\vert\uparrow\uparrow\rangle_{12}+e^{-i(\frac{J_z}{4}-h')t}(\vert\uparrow\downarrow\rangle_{12}+\vert\downarrow\uparrow\rangle_{12})+e^{-i(\frac{J_z}{4}+J_0-h-h')t}\vert\downarrow\downarrow\rangle_{12}     \right].\nonumber\\
\label{purestateof12spins}
\end{eqnarray}}
Let us study the entanglement of $S_1$ and $S_2$ spins. For this purpose, we average the density matrix of the system $\rho(t)=\vert\psi(t)\rangle\langle\psi(t)\vert$ over the state of the $S_a$ and $S_b$ spins
\begin{eqnarray}
\rho(t)_{12}={\rm Tr}_{ab}\rho(t)=\frac{1}{4}\left(\vert\phi_1\rangle_{12}\langle\phi_1\vert_{12}+2\vert\phi_2\rangle_{12}\langle\phi_2\vert_{12}+\vert\phi_3\rangle_{12}\langle\phi_3\vert_{12}\right).
\label{spin12densmatrix}
\end{eqnarray}
As we can see the mixed state of $S_1$ and $S_2$ spins consist of the ensemble of pure states which define by formula (\ref{purestateof12spins}). Using the definition of the entanglement of formation (\ref{entangledofformation}),
we easily calculate the value of entanglement between spins $S_1$ and $S_2$ in state (\ref{spin12densmatrix}). For this purpose, entropy of entanglement (\ref{entropyofentanglement}) of states $\vert\phi_i\rangle$ (\ref{purestateof12spins}),
should be calculated. Since the entropy of entanglement of each of the states (\ref{purestateof12spins}) is equal to zero, then the value of entanglement of formation between spins $S_1$ and $S_2$ in state (\ref{spin12densmatrix})
is equal zero $E(\rho_{12}(t))=0$ (see Appendix \ref{appb}). The state (\ref{spin12densmatrix}) is written in decomposition with probabilities that correspond to the minimum possible value
in expression (\ref{entangledofformation}).

\subsection{Entanglement between subsystems $S_a$, $S_b$ and $S_1$, $S_2$ \label{entbetweentwosubsys}}

\begin{figure}[!!h]
\includegraphics[scale=1.00, angle=0.0, clip]{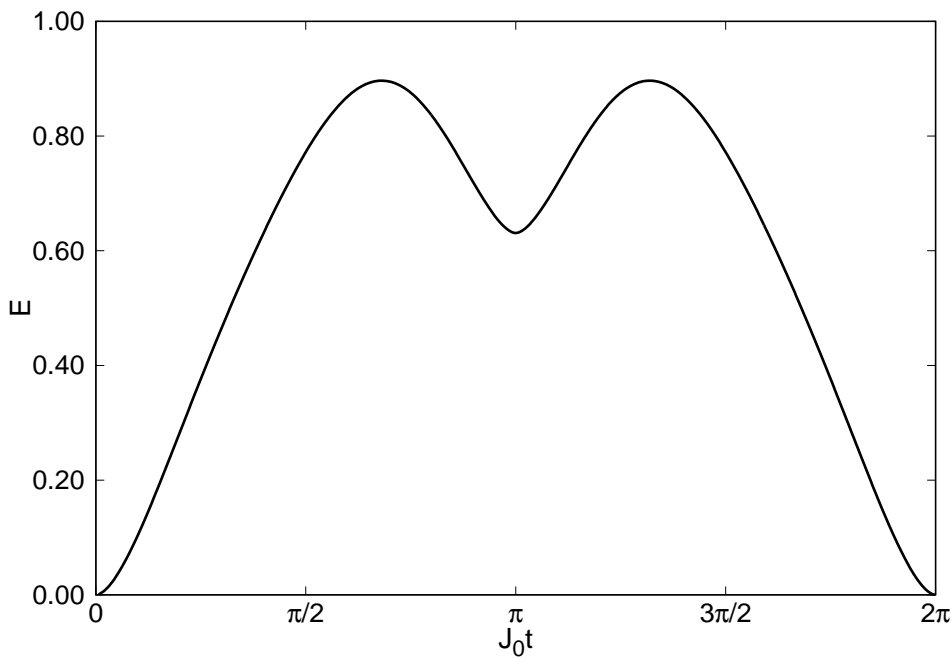}
\caption{The time dependence of entropy of entanglement (\ref{entropyofentanglementsubsystem}) between subsystems consisting of $S_a$, $S_b$ spins and $S_1$, $S_2$ spins in state (\ref{evolution12ab}).
In numerical calculations, the logarithm to base 3 is taken.}
\label{ent_subsyst_ab_12}
\end{figure}

Now let us study the entanglement between subsystems consisting of two central spins $S_a$, $S_b$ and two side spins $S_1$, $S_2$ spins. Due to the fact that the state of the entire system is pure (\ref{evolution12ab}), we use
the definition of von Neumann entropy (\ref{entropyofentanglement}) to calculate the entanglement. We calculate the entropy of the subsystem consisting of $S_a$, $S_b$ spins. For this purpose,
the density matrix of spins $S_a$, $S_b$ (\ref{appa11}) should be substituted in definition (\ref{entropyofentanglement}). As a result, we obtain expression describing the entanglement
between subsystems in the form
\begin{eqnarray}
E(\vert\psi(t)\rangle)=-{\rm Tr}\left(\rho_{ab}(t)\log\rho_{ab}(t)\right)=-\sum_{\lambda_i}\lambda_i\log\lambda_i,
\label{entropyofentanglementsubsystem}
\end{eqnarray}
where $\lambda_i$ are the eigenvalues of density matrix $\rho_{ab}(t)$ (\ref{appa11}). From the eigenvalue equation $\det\vert\rho_{ab}(t)-\lambda I\vert=0$ we obtain the following cubic equation
\begin{eqnarray}
\left(\frac{1}{4}-\lambda\right)^3+\frac{1}{4}\left(\frac{1}{4}-\lambda\right)^2-\frac{1}{16}\left(B^4+(1+B)^2\right)\left(\frac{1}{4}-\lambda\right)+\frac{1}{64}B^2\left(1+2B\right)=0.\nonumber\\
\label{cubicequationenta}
\end{eqnarray}
Solving this equation numerically with respect to $\lambda$ and substituting solutions into the formula (\ref{entropyofentanglementsubsystem}), we obtain the value of entanglement between subsystems $S_a$, $S_b$ and $S_1$, $S_2$
as a function of time (Fig.~\ref{ent_subsyst_ab_12}).
It is easy to see that entanglement is periodic with period $2\pi$ with respect to the time parameter $J_0t$. It has also two maxima $E_{max}\approx 0.8966$ at moments of time $J_0t_{max}\approx 2.14$ and $4.14$
and one local minimum $E_{min}\approx 0.6310$ at the moment of time $J_0t_{min}=\pi$ which are periodic with period $2\pi$. These points can be directly found from the array of numerical calculations
of entropy $E$. It is also worth noting that in this case, the entanglement does not achieve the maximum possible value $E=1$.
This can be seen from the form of density matrix (\ref{appa11}). In order, to achieve the maximum entanglement, it is necessary that the density matrix be a unit operator. However, there is no such time
moment when it becomes a unit operator.

\subsection{Entanglement between one spin and the rest of the system \label{entaspin1withother}}

In this subsection, we calculate the entanglement between one spin and the rest of the system. The system has symmetry with respect to $S_a$ and $S_b$ spins and to $S_1$ and $S_2$ spins.
This means that the behaviour of entanglement of $S_a$ spin with the rest of system is the similar to the bevaiour of entanglement of $S_b$ spin with the rest of the system. We have the same situation
in the case of $S_1$, $S_2$ spins. Therefore it is enough to calculate the entanglement of spin $S_a$ with the rest of the system and the entanglement of spin $S_1$ with the rest of the system.
To calculate entanglement in the first case the density matrix of the $S_a$ spin in the state (\ref{evolution12ab}) should be obtained. For this purpose, we average density matrix (\ref{spinabdensmatrix})
over the state of spin $S_b$. It takes the form
\begin{eqnarray}
&&\rho(t)_{a}={\rm Tr}_{b}\rho(t)_{ab}\nonumber\\
&&=\frac{1}{2}\left(\vert\uparrow\rangle_{a}\langle\uparrow\vert_{a}+\cos\frac{(J_z-J)t}{2}\cos^2\frac{J_0t}{2}\left(e^{-ih't}\vert\uparrow\rangle_{a}\langle\downarrow\vert_{a}+e^{ih't}\vert\downarrow\rangle_{a}\langle\uparrow\vert_{a}\right)+\vert\downarrow\rangle_{a}\langle\downarrow\vert_{a}\right).\nonumber\\
\label{densmatrixa}
\end{eqnarray}
In this case, the von Neumann entanglement entropy (\ref{entropyofentanglement}) has the form
\begin{eqnarray}
E(\vert\psi\rangle)=-{\rm Tr}\left(\rho_a\log\rho_a\right)=-\lambda_+\log\lambda_+-\lambda_-\log\lambda_-,
\label{entropyofentanglementcasespina}
\end{eqnarray}
where
\begin{eqnarray}
\lambda_{\pm}=\frac{1}{2}\left(1\pm\cos\frac{(J_z-J)t}{2}\cos^2\frac{J_0t}{2}\right)
\label{eigenvalueofrhoa}
\end{eqnarray}
are the eigenvalues of density matrix (\ref{densmatrixa}). As we can see that entanglement of the $S_a$ or $S_b$ spin with the rest system depends on parameters $J_z-J$ and $J_0$ and takes the maximum value
when $\cos\frac{(J_z-J)t}{2}\cos^2\frac{J_0t}{2}=0$. Then the density matrix of the $S_a$ or $S_b$ spin (\ref{densmatrixa}) becomes a unit operator.

To obtain the behaviour of entanglement between $S_1$ spin and the rest system
we average density matrix (\ref{spin12densmatrix}) over the state of spin $S_2$. It takes the form
\begin{eqnarray}
&&\rho(t)_{1}={\rm Tr}_{2}\rho(t)_{12}\nonumber\\
&&=\frac{1}{2}\left(\vert\uparrow\rangle_{1}\langle\uparrow\vert_{1}+\cos^2\frac{J_0t}{2}\left(e^{-ih't}\vert\uparrow\rangle_{1}\langle\downarrow\vert_{1}+e^{ih't}\vert\downarrow\rangle_{1}\langle\uparrow\vert_{1}\right)+\vert\downarrow\rangle_{1}\langle\downarrow\vert_{1}\right).\nonumber\\
\label{densmatrix1}
\end{eqnarray}
Naturally, this density matrix depends only on the parameter $J_0$. Because this spin interacts with the rest of the system through this parameter. The entanglement between $S_1$ or $S_2$ spins and the rest system
is defined by equation (\ref{entropyofentanglementcasespina}) with
\begin{eqnarray}
\lambda_{\pm}=\frac{1}{2}\left(1\pm\cos^2\frac{J_0t}{2}\right).
\label{eigenvalueofrhoa}
\end{eqnarray}
It is clear that the entanglement takes the maximal value when $\cos^2\frac{J_0t}{2}=0$.

\section{Conclusions \label{conc}}

We have studied the evolution of entanglement in the diamond spin-$1/2$ cluster. The cluster consists of two central spins which describe by the anisotropic Heisenberg model and connect with two side spins via Ising interaction.
Using the Wootters definition of concurrence, the dynamic of entanglement between central spins has been calculated. The entanglement between these spins depends on the difference between the interaction couplings
and the influence of the side spins. It has been shown the greater difference between the interaction couplings of the central spins, the faster the entanglement grows in the system.
We have examined the influence of the coupling constant of the side spins on the entanglement of the central spins. As a result, we have obtained that the interaction with the side spins reduces the entanglement between central spins.
We have also studied the behaviour of the entanglement depending on the ratio between the coupling constants of central and side spins. The magnitude of this ratio affects both the values of entanglement achieved during evolution
and its periodicity. Depending on the ratio between interacting couplings, the conditions for achieving minimal and maximal values of entanglement between central spins (\ref{minpointofent}),
(\ref{maxpointofent}) have been obtained. Thereby, the side spins allow one to control the behaviour of entanglement between the central spins, which is important for the implementation of various algorithms of quantum information
and for providing quantum calculations on such system.

The entanglement between side spins has been also investigated. We have shown that during the evolution of the system, it equals zero. Finally, we have studied the bipartite entanglement between different parts of the system and
have obtained conditions for achieving maximally entanglement states. We have studied the evolution of entanglement between subsystems of central spins and side spins. Due to the
nature of the interaction between spins, there is no such time moment when the density matrix (\ref{appa11}) becomes a unit operator. Therefore the entanglement between these subsystems never takes
the maximal possible values. Finally, we have obtained the analytical expressions to determine the entanglement of one central spin and the rest of the system (\ref{entropyofentanglementcasespina}), (\ref{eigenvalueofrhoa})
and one side spin with the rest of the system (\ref{entropyofentanglementcasespina}), (\ref{eigenvalueofrhoa}).

\section{Acknowledgements}

This work was supported by Project FF-27F (No.~0122U001558) from the Ministry of Education and Science of Ukraine.

\begin{appendices}

\section{Entanglement of state $\rho_{ab}(t)$ \label{appa}}
\setcounter{equation}{0}
\renewcommand{\theequation}{A\arabic{equation}}

Using Wootters definition of measurement of entanglement (\ref{wootters}), we calculate the value of entanglement between spins $S_a$ and $S_b$ in state (\ref{appa11}).
The matrix $\tilde{\rho}(t)_{ab}$ takes the form
\begin{eqnarray}
&&\tilde{\rho}(t)_{ab}= \sigma^y\otimes\sigma^y \rho^*(t)_{ab}\sigma^y\otimes\sigma^y\nonumber\\
&&=\left( \begin{array}{ccccc}
\frac{1}{4} & -\frac{1}{8}e^{-ih't}A^*(1+B) & -\frac{1}{8}e^{-ih't}A^*(1+B) & \frac{1}{4}e^{-i2h't}B^2\\[9pt]
-\frac{1}{8}e^{ih't}A(1+B) & \frac{1}{4} & \frac{1}{4} & -\frac{1}{8}e^{-ih't}A(1+B) \\[9pt]
-\frac{1}{8}e^{ih't}A(1+B) & \frac{1}{4} & \frac{1}{4} & -\frac{1}{8}e^{-ih't}A(1+B) \\[9pt]
\frac{1}{4}e^{i2h't}B^2 & -\frac{1}{8}e^{ih't}A^*(1+B) & -\frac{1}{8}e^{ih't}A^*(1+B) & \frac{1}{4}
\end{array}\right).\nonumber\\
\label{appa12}
\end{eqnarray}
Using equations (\ref{appa11}) and (\ref{appa12}), we construct the following matrix
\begin{eqnarray}
&&\rho(t)_{ab}\tilde{\rho}(t)_{ab}\nonumber\\
&&=\left( \begin{array}{ccccc}
\frac{1}{16}\left(1+B^4\right)-\frac{1}{32}A^2\left(1+B\right)^2 & -\frac{1}{32}e^{-ih't}\left(1+B\right)\left(A^*\left(1+B^2\right)-2A\right) \\[9pt]
\frac{1}{32}e^{ih't}\left(1+B\right)\left(A^*\left(1+B^2\right)-2A\right) & -\frac{1}{32}{A^*}^2\left(1+B\right)^2+\frac{1}{8}  \\[9pt]
\frac{1}{32}e^{ih't}\left(1+B\right)\left(A^*\left(1+B^2\right)-2A\right) & -\frac{1}{32}{A^*}^2\left(1+B\right)^2+\frac{1}{8}  \\[9pt]
\frac{1}{8}e^{i2h't}B^2-\frac{1}{32}e^{i2h't}A^2\left(1+B\right)^2 & -\frac{1}{32}e^{ih't}\left(1+B\right)\left(A^*\left(1+B^2\right)-2A\right)
\end{array}\right.\nonumber
\end{eqnarray}

\begin{eqnarray}
\left. \begin{array}{ccccc}
-\frac{1}{32}e^{-ih't}\left(1+B\right)\left(A^*\left(1+B^2\right)-2A\right) & \frac{1}{8}e^{-i2h't}B^2-\frac{1}{32}e^{-i2h't}A^2\left(1+B\right)^2 \\[9pt]
-\frac{1}{32}{A^*}^2\left(1+B\right)^2+\frac{1}{8} & \frac{1}{32}e^{-ih't}\left(1+B\right)\left(A^*\left(1+B^2\right)-2A\right) \\[9pt]
-\frac{1}{32}{A^*}^2\left(1+B\right)^2+\frac{1}{8} & \frac{1}{32}e^{-ih't}\left(1+B\right)\left(A^*\left(1+B^2\right)-2A\right) \\[9pt]
-\frac{1}{32}e^{ih't}\left(1+B\right)\left(A^*\left(1+B^2\right)-2A\right) & \frac{1}{16}\left(1+B^4\right)-\frac{1}{32}A^2\left(1+B\right)^2
\end{array} \right).
\label{appa13}
\end{eqnarray}
From the eigenvalue equation $\det\vert\rho(t)_{ab}\tilde{\rho}(t)_{ab}-\omega I\vert=0$, we obtain the following equations for $\omega$
\begin{eqnarray}
&&\omega^2=0,\quad \omega^2=\frac{1}{16}\left(1-B^2\right)^2,\nonumber\\
&&\omega^4+\frac{1}{16}\left((A^2+{A^*}^2)(1+B)^2-(1+B^2)^2-4\right)\omega^2+\frac{1}{16^2}\left(1-B\right)^4=0.\nonumber\\
\label{appa14}
\end{eqnarray}
Solving these equations with respect to $\omega$ and taking into formula (\ref{wootters}) only positive solutions we obtain the concurrence as a function of time between spins $S_a$ and $S_b$.

\section{Entanglement of state $\rho_{12}(t)$ \label{appb}}
\setcounter{equation}{0}
\renewcommand{\theequation}{A\arabic{equation}}

Let us calculate the value of entanglement between spins $S_1$ and $S_2$ in state (\ref{spin12densmatrix}). For this purpose, we calculate the entropy of entanglement (\ref{entropyofentanglement})
of pure states (\ref{purestateof12spins}) which form mixed state $\rho_{12}(t)$. Thus, for state $\vert\phi_i\rangle_{12}$ the entropy of entanglement is defined by equation
\begin{eqnarray}
E(\vert\phi_i\rangle_{12})=-{\rm Tr}\left(\rho_1\log\rho_1\right)=-{\rm Tr}\left(\rho_2\log\rho_2\right),
\label{appaeq1}
\end{eqnarray}
where $\rho_{1(2)}={\rm Tr}_{2(1)}\vert\phi_i\rangle_{12}\langle\phi_i\vert_{12}$ is the partial trace of the state over the subsystem $2(1)$. It is easy to calculate the density matrices of one spin in states (\ref{purestateof12spins}).
For state $\vert\phi_1\rangle_{12}$ the density matrix of spin $S_1$ takes the form
\begin{eqnarray}
&&\rho_1={\rm Tr}_2 \vert\phi_1\rangle_{12}\langle\phi_1\vert_{12}\nonumber\\
&& = \frac{1}{2}\left(\vert\uparrow\rangle_1\langle\uparrow\vert_1+e^{-i(J_0+h)t}\vert\uparrow\rangle_1\langle\downarrow\vert_1 +e^{i(J_0+h)t}\vert\downarrow\rangle_1\langle\uparrow\vert_1 +\vert\downarrow\rangle_1\langle\downarrow\vert_1 \right)\nonumber\\
\label{appaeq2}
\end{eqnarray}
To calculate the entropy of this matrix, we find its eigenvalues. As a result, we obtain the following equation for eigenvalues
\begin{eqnarray}
\left(\frac{1}{2}-\lambda\right)^2=\frac{1}{4}.
\label{appaeq3}
\end{eqnarray}
It has two roots $\lambda_+=0$ and $\lambda_-=1$. Substituting these roots in formula (\ref{appaeq1}), we obtain
\begin{eqnarray}
E(\vert\phi_1\rangle_{12})=-{\rm Tr}\left(\rho_1\log\rho_1\right)=-\lambda_+\log\lambda_+-\lambda_-\log\lambda_-=0.
\label{appaeq4}
\end{eqnarray}
In the similar way, for state $\vert\phi_2\rangle_{12}$ and $\vert\phi_3\rangle_{12}$, we arrive the same equation for the eigenvalues (\ref{appaeq3}) as in the case of state $\vert\phi_1\rangle_{12}$.
Accordingly entropy of entanglement for states $\vert\phi_2\rangle_{12}$ and $\vert\phi_3\rangle_{12}$ are equal to zero
\begin{eqnarray}
E(\vert\phi_2\rangle_{12})=E(\vert\phi_3\rangle_{12})=0.
\label{appaeq5}
\end{eqnarray}
Now we put the values of entanglement (\ref{appaeq4}), (\ref{appaeq5}) in formula (\ref{entangledofformation}) and obtain the value of entanglement of formation between spins $S_1$ and $S_2$ in state (\ref{spin12densmatrix})
\begin{eqnarray}
E(\rho_{12}(t))=\min\left(p_1 E(\vert\phi_1\rangle_{12})+p_2 E(\vert\phi_2\rangle_{12})+p_3 E(\vert\phi_3\rangle_{12})\right)=0.
\label{appaeq6}
\end{eqnarray}

\end{appendices}

{}


\begin{thebibliography}{99}
\bibitem{einstein1935} A. Einstein, B. Podolsky, N. Rosen, Phys. Rev. {\bf 47}, 777 (1935).
\bibitem{bell1964} J. S. Bell, Physics {\bf 1}, 195 (1964).
\bibitem{aspect982} A. Aspect, P. Grangier, G. Roger, Phys. Rev. Lett. {\bf 49}, 91 (1982).
\bibitem{Ekert1991} A. K. Ekert, Phys. Rev. Lett. {\bf 67}, 661 (1991).
\bibitem{Bennett1992} Ch. H. Bennett, S. J. Wiesner, Phys. Rev. Lett. {\bf 69}, 2881 (1992).
\bibitem{TELEPORT} C. H. Bennett, G. Brassard, C. Crepeau, R. Jozsa, A. Peres, W. K. Wootters, Phys. Rev. Lett. {\bf 70}, 1895 (1993).
\bibitem{Zeilinger1997} D. Bouwmeester, J.-W. Pan, K. Mattle, M. Eibl, H. Weinfurter, A. Zeilinger, Nature {\bf 390}, 575 (1997).
\bibitem{cerf1998} N. J. Cerf, C. Adami, P. G. Kwiat, Phys. Rev. A {\bf 57}, R1477 (1998).
\bibitem{pittman2001} T. B. Pittman, B. C. Jacobs, J. D. Franson, Phys. Rev. A {\bf 64}, 062311 (2001).
\bibitem{gasparoni2004} S. Gasparoni, J.-W. Pan, Ph. Walther, T. Rudolph, A. Zeilinger, Phys. Rev. Lett. {\bf 93}, 020504 (2004).
\bibitem{Giovannetti20031} V. Giovannetti, S. Lloyd and L. Maccone, Europhys. Lett. {\bf 62}, 615 (2003).
\bibitem{Giovannetti20032} V. Giovannetti, S. Lloyd and L. Maccone, Phys. Rev. A {\bf 67}, 052109 (2003).
\bibitem{Batle2005} J. Batle, M. Casas, A. Plastino and A. R. Plastino, Phys. Rev. A {\bf 72}, 032337 (2005).
\bibitem{Borras2006} A. Borras, M. Casas, A. R. Plastino and A. Plastino, Phys. Rev. A {\bf 74}, 022326 (2006).
\bibitem{ASPECT} A. Aspect, J. Dalibard, G. Roger, Phys. Rev. Lett. {\bf 49}, 1804 (1982).
\bibitem{quantcomp} David P. DiVincenzo, Fortschr. Phys. \textbf{48}, 771 (2000).
\bibitem{qdots1} Daniel Loss and David P. DiVincenzo, Phys. Rev. A \textbf{57}, 120 (1998).
\bibitem{phosphorus3} B. E. Kane, Nature \textbf{393}, 133 (1998).
\bibitem{phosphorus1} Jarryd J. Pla, Kuan Y. Tan, Juan P. Dehollain, Wee H. Lim, John J. L. Morton, Floris A. Zwanenburg, David N. Jamieson, Andrew S. Dzurak and Andrea Morello, Nature \textbf{496}, 334 (2013).
\bibitem{kuzmak2020} A. R. Kuzmak, Phys. Scr. {\bf 95}, 035403 (2020).
\bibitem{supcond1} L. F. Wei, Yu-xi Liu and Franco Nori, Phys. Rev. B \textbf{71}, 134506 (2005).
\bibitem{supcond2} J. E. Mooij, T. P. Orlando, L. Levitov, Lin Tian, Caspar H. van der Wal  and Seth Lloyd, Science \textbf{285}, 1036 (1999).
\bibitem{supcond3} Yuriy Makhlin, Gerd Sch\"on and Alexander Shnirman, Rev. Mod. Phys. \textbf{73}, 357 (2001).
\bibitem{supcond4} J. Majer et al., Nature \textbf{449}, 443 (2007).
\bibitem{drillon1988} M. Drillon, E. Coronado, M. Belaiche, R. L. Carlin, J. Appl. Phys. {\bf 63}, 3551 (1988).
\bibitem{drillon1993} M. Drillon, M. Belaiche, P. Legoll, J. Aride, A. Boukhari, A. Moqine, J. Magn. Magn. Mater. {\bf 128}, 83 (1993).
\bibitem{sakurai2002} H. Sakurai, K. Yoshimura, K. Kosuge, N. Tsujii, H. Abe, H. Kitazawa, G. Kido, H. Michor, G. Hilscher, J. Phys. Soc. Japan {\bf 71}, 1161 (2002).
\bibitem{kikuchi2005} H. Kikuchi, Y. Fujii, M. Chiba, S. Mitsudo, T. Klehara, T. Tonegawa, K. Okamoto, T. Sakai, T. Kuwai, H. Ohta, Phys. Rev. Lett. {\bf 94}, 227201 (2005).
\bibitem{ananikian2012} N. Ananikian, H. Lazaryan, M. Nalbandyan, Eur. Phys. J. B {\bf 85}, 223 (2012).
\bibitem{valverde2008} J. S. Valverde, O. Rojas, S. M. de Souza, J. Phys.: Condens. Matter {\bf 20}, 345208 (2008).
\bibitem{Ishii2000} M. Ishii, H. Tanaka, M. Hori, H. Uekusa, Y. Ohashi, K. Tatani, Y. Narumi, K. Kindo, J. Phys. Soc. Jpn. {\bf 69}, 340 (2000).
\bibitem{honecker2001} A. Honecker, A. Läuchli, Phys. Rev. B {\bf 63}, 174407 (2001).
\bibitem{canova2006} L. Canova, J. Strecka, M. Jascur, J. Phys.: Condens. Matter {\bf 18}, 4967 (2006).
\bibitem{gu2007} Bo Gu, Gang Su, Phys. Rev. B {\bf 75}, 174437 (2007).
\bibitem{carvalho2019} I. M. Carvalho, J. Torrico, S. M. de Souza, O. Rojas, O. Derzhko, Ann. Phys. (NY) {\bf 402}, 45 (2019).
\bibitem{bose2005} I. Bose, A. Tribedi, Phys. Rev. A {\bf 72}, 022314 (2005).
\bibitem{tribedi2006} A. Tribedi, S. Bose, Phys. Rev. A {\bf 74}, 012314 (2006).
\bibitem{ananikian2006} N. S. Ananikian, L. N. Ananikyan, L. A. Chakhmakhchyan, O. Rojas, J. Phys.: Condens. Matter {\bf 24}, 256001 (2012).
\bibitem{chakhmakhchyan2012} L. Chakhmakhchyan,, N. Ananikian, L. Ananikyan, C. Burdik, J. Phys.: Conf. Ser. {\bf 343}, 012022 (2012).
\bibitem{rojas2012} O. Rojas, M. Rojas, N. S. Ananikian, S. M. de Souza, Phys. Rev. A {\bf 86}, 042330 (2012).
\bibitem{rojas2014} J. Torrico, M. Rojas, S. M. de Souza, O. Rojas, N. S. Ananikian, EPL {\bf 108}, 50007 (2014).
\bibitem{torrico2016} J. Torrico, M. Rojas, M. S. S. Pereira, J. Strecka, M. L. Lyra, Phys. Rev. B {\bf 93}, 014428 (2016).
\bibitem{rojas2017} O. Rojas, M. Rojas, S. M. de Souza, J. Torrico, J. Strecka, M. L. Lyra, Physica A {\bf 486}, 367 (2017).
\bibitem{Zheng2018} Y. Zheng, Z. Mao, B. Zhou, Chin. Phys. B {\bf 27}, 090306 (2018).
\bibitem{Cavalho2019} I. M. Carvalho, O. Rojas, S. M. de Souza, M. Rojas, Quant. Inf. Process. {\bf 18}, 134 (2019).
\bibitem{Ghannadan2022} A. Ghannadan, Katarína Karl'ova, J. Strecka, Magnetochemistry {\bf 8}, 11 (2022).
\bibitem{escuer1998} A. Escuer, R. Vicente, S. B. Kumar, F. A. Mautner, J. Chem. Soc. Dalton Trans. {\bf 20}, 3473 (1998).
\bibitem{hagiwara2006} M. Hagiwara, Y. Narumi, A. Matsuo, H. Yashiro, S. Kimura, K. Kundo, New J. Phys. {\bf 8}, 176 (2006).
\bibitem{desurvire2009} E. Desurvire, {\it Classical and Quantum Information Theory: An Introduction for the Telecom Scientist} (Cambridge University Press, Cambridge, 2009).
\bibitem{wootters1998} W. K. Wootters, Phys. Rev. Lett. {\bf 80}, 2245 (1998).
\bibitem{wootters1997} S. A. Hill, W. K. Wootters, Phys. Rev. Lett. {\bf 78}, 5022 (1997).
\bibitem{wootters19960} C. H. Bennett, H. J. Bernstein, S. Popescu, B. Schumacher, Phys. Rev. A {\bf 53}, 2046 (1996)
\bibitem{popescu1997} S. Popescu, D. Rohrlich, Phys. Rev. A {\bf 56}, R3319(R) (1997).
\bibitem{bennett1996} C. H. Bennett, D. P. DiVincenzo, J. A. Smolin, W. K. Wootters, Phys. Rev. A {\bf 54}, 3824 (1996).
\end{thebibliography}
\end{document}